\documentclass[12pt,a4paper]{article} 
\pdfoutput=1
\usepackage{graphicx}
\usepackage{etex}
\usepackage{jheppub}
\usepackage{epstopdf}
\usepackage{epsfig}
\usepackage{dcolumn}  
\usepackage{bm}    
\usepackage{amssymb} 
\usepackage{amsmath,bm}
\usepackage{amsfonts}  
\usepackage{slashed}  
\usepackage{youngtab}
\usepackage[mathscr]{euscript}
\usepackage{epsfig}
\usepackage[utf8]{inputenc}
\usepackage{tikz}
\usetikzlibrary{tikzmark,calc,,arrows,shapes,decorations.pathreplacing}
\usepackage{xcolor}
\usepackage{mathrsfs}
\usepackage{verbatim}
\usepackage{cases}
\usepackage{bbding}
\usepackage{pifont}
\usepackage{ulem}
\usepackage{dcolumn}
\usepackage{graphicx}
\usepackage{pgf}
\usepackage{pstricks}
\usepackage{pst-node}
\usepackage[all]{xy}
\usepackage[figuresright]{rotating}
\usepackage{hyperref}
\usepackage{url}
\usepackage{titlesec}

\titleclass{\subsubsubsection}{straight}[\subsection]

\newcounter{subsubsubsection}[subsubsection]
\renewcommand\thesubsubsubsection{\thesubsubsection.\arabic{subsubsubsection}}
\titleformat{\subsubsubsection}
  {\normalfont\normalsize\bfseries}{\thesubsubsubsection}{1em}{}
\titlespacing*{\subsubsubsection}
{0pt}{3.25ex plus 1ex minus .2ex}{1.5ex plus .2ex}

\makeatletter
\renewcommand\paragraph{\@startsection{paragraph}{5}{\z@}%
  {3.25ex \@plus1ex \@minus.2ex}%
  {-1em}%
  {\normalfont\normalsize\bfseries}}
\renewcommand\subparagraph{\@startsection{subparagraph}{6}{\parindent}%
  {3.25ex \@plus1ex \@minus .2ex}%
  {-1em}%
  {\normalfont\normalsize\bfseries}}
\def\toclevel@subsubsubsection{4}
\def\toclevel@paragraph{5}
\def\toclevel@paragraph{6}
\def\l@subsubsubsection{\@dottedtocline{4}{7em}{4em}}
\def\l@paragraph{\@dottedtocline{5}{10em}{5em}}
\def\l@subparagraph{\@dottedtocline{6}{14em}{6em}}
\makeatother

\usepackage{young}
\usepackage{graphicx,rotating,booktabs}

\usepackage{varwidth}

\newdimen\tableauside\tableauside=1.0ex
\newdimen\tableaurule\tableaurule=0.4pt
\newdimen\tableaustep
\def\phantomhrule#1{\hbox{\vbox to0pt{\hrule height\tableaurule width#1\vss}}}
\def\phantomvrule#1{\vbox{\hbox to0pt{\vrule width\tableaurule height#1\hss}}}
\def\sqr{\vbox{%
		\phantomhrule\tableaustep
		\hbox{\phantomvrule\tableaustep\kern\tableaustep\phantomvrule\tableaustep}%
		\hbox{\vbox{\phantomhrule\tableauside}\kern-\tableaurule}}}
\def\squares#1{\hbox{\count0=#1\noindent\loop\sqr
		\advance\count0 by-1 \ifnum\count0>0\repeat}}
\def\tableau#1{\vcenter{\offinterlineskip
		\tableaustep=\tableauside\advance\tableaustep by-\tableaurule
		\kern\normallineskip\hbox
		{\kern\normallineskip\vbox
			{\gettableau#1 0 }%
			\kern\normallineskip\kern\tableaurule}%
		\kern\normallineskip\kern\tableaurule}}
\def\gettableau#1 {\ifnum#1=0\let\next=\null\else
	{{\tiny\yng(1)}}s{#1}\let\next=\gettableau\fi\next}

\def\ASF{$\textrm{AdS}_5$$\times$$S^5$}
\def\AST{$\textrm{AdS}_3$$\times$$S^3$}
\def\ASTT{$\textrm{AdS}_3$$\times$$S^3$$\times$$T^4$}
\def\ASTM{$\textrm{AdS}_3$$\times$$S^3$$\times$$M_4$}

\def\IFsp{\mathsf{I}_{\textrm{s.p.}}}
\def\IFf{\mathcal{I}_{\textrm{full}}}
\def\IFsl{\mathfrak{i}}

\def\pfWS{\mathscr{Z}}
\def\IWS{\widetilde{\mathscr{Z}}}
\def\pfWSsl{\mathfrak{z}}
\def\pfWSw{\mathfrak{z}}
\def\IWSsl{\tilde{\mathfrak{z}}}
\def\IWSw{\tilde{\mathfrak{z}}}

\def\ChNf{\chi^{\mathcal{N}=4}}

\def\ChShNf{\mathsf{X}^{\mathcal{N}=4,\textrm{ short}}}
\def\ChLoNf{\mathsf{X}^{\mathcal{N}=4,\textrm{ long}}}
\def\IShNf{\widetilde{\mathsf{X}}^{\mathcal{N}=4,\textrm{ short}}}
\def\ILoNf{\widetilde{\mathsf{X}}^{\mathcal{N}=4,\textrm{ long}}}

\def\myng{\tiny\yng}

\def\myq{\mathfrak{q}}

\newcommand{\includeCroppedPdf}[2][]{%
    \IfFileExists{./#2-crop.pdf}{}{%
        \immediate\write18{pdfcrop #2 #2-crop.pdf}}%
    \includegraphics[#1]{#2-crop.pdf}}

\newcommand{\be}{ \begin{equation}}
\newcommand{\ee}{\end{equation}}
\newcommand{\bea}[1]{\begin{eqnarray}\label{#1} }
\newcommand{\eea}{\end{eqnarray}}

\def\ZZZ{{\hskip-3pt\hbox{ Z\kern-1.6mm Z}}}
\def\zzz{{\hskip-3pt\hbox{ z\kern-1mm z}}}

\def\bal#1\eal{\begin{align}#1\end{align}}

\def\PE{\textrm{PE}}
\def\Plog{\textrm{Plog}}

\title{Chiral algebra from worldsheet}
\author{Wei Li}
\affiliation{
Institute of Theoretical Physics, Chinese Academy of Sciences,\\
\hspace*{0.3cm}100190 Beijing, P.R.\ China
} 
\emailAdd{weili@mail.itp.ac.cn}

\date{today}

\abstract{
The chiral algebra of a 4D $\mathcal{N}\geq2$ superconformal field theory is a vertex operator algebra generated by the Schur subsector of the 4D theory and its rigid (yet rich) structure has been useful in constraining and classifying 4D $\mathcal{N}=2$ SCFTs.
We study how the chiral algebra arises from the worldsheet perspective. 
In the worldsheet CFT dual of 4D $\mathcal{N}=4$ SYM at the free point, we extract the subsector that corresponds to the spacetime Schur operators at \textit{generic} coupling, and show how they generate the chiral algebra.
The result can be easily generalized to 4D $\mathcal{N}=2$ superconformal field theories that arise as orbifolds of 4D $\mathcal{N}=4$ SYM. 
}

\begin{document}
\maketitle

\section{Introduction}

For a general 4D $\mathcal{N}\geq2$ superconformal field theory (SCFT), there exists a (2D) vertex operator algebra (called chiral algebra) structure underlying its Schur subsector \cite{Beem:2013sza}. 
It can be defined by considering the BRST cohomology of the linear combination of a Poincar\'{e} supercharge and a conformal one, which selects the Schur operators  restricted to $\mathbb{R}^2\subset \mathbb{R}^4$; the correlation functions of these local operators are meromorphic functions of the $\mathbb{R}^2$ coordinates and are determined by the chiral algebra.
The chiral algebra is infinitely dimensional,  and its rigid but rich structure has proven to be useful in constraining, bootstrapping, and classifying the 4D theories, see e.g.\ \cite{Beem:2014rza,Beem:2014zpa,Beem:2016wfs}.\footnote{The story also has a generalization to 6D $(2,0)$ SCFTs \cite{Beem:2014kka,Beem:2015aoa}.}

The chiral algebra can also be viewed as the algebra generated by the local operators of the holomorphic-topological twist (also called Kapustin twist \cite{Kapustin:2006hi}) of the 4D theory on $\mathbb{C} \times \Sigma$, in the presence of $\Omega$-background \cite{Costello:2018fnz,Oh:2019bgz,Jeong:2019pzg,Butson:2020mmu}. 
In accordance with this perspective, holographically, the chiral algebra should arise from twisting the bulk supergravity, as in
\cite{Costello:2016mgj}. 
Indeed, for 4D $\mathcal{N}=4$ SYM, its chiral algebra was derived from the topological B-model of the deformed conifold, which is the result of twisting IIB supergravity \cite{Costello:2018zrm}.
For further developments and generalizations see e.g.\ \cite{Budzik:2021fyh,Budzik:2022hcd,Costello:2023hmi}.

\medskip

The goal of this paper is to understand the chiral algebra from a string worldsheet perspective. 
In particular, we start with the worldsheet CFT dual of free 4D $\mathcal{N}=4$ SYM of \cite{Gaberdiel:2021qbb, Gaberdiel:2021jrv} and derive the chiral algebra directly from the worldsheet CFT.
This provides an alternative derivation of the chiral algebra to	 the one from field theory or gravity,  

\medskip

Apart from providing another perspective on the chiral algebra, one motivation for this work is to better understand the worldsheet CFT dual of tensionless strings in AdS$_5\times$S$^5$. 
Finding the string worldsheet theory for AdS$_5\times$S$^5$ has been a difficult problem.
Recently, building on the success of the worldsheet theory for tensionless strings in AdS$_3\times$S$^3\times$M$_4$ \cite{Eberhardt:2018ouy,Eberhardt:2019niq}, with M$_4$ being $T^4$, K3, or S$^3\times$S$^1$, a worldsheet theory was proposed for tensionless strings in AdS$_5\times$S$^5$, and was shown to correctly reproduce the spectrum of 4D $\mathcal{N}=4$ free SYM at large-$N$  \cite{Gaberdiel:2021qbb, Gaberdiel:2021jrv}.

This worldsheet theory for AdS$_5\times$S$^5$ relies on a conjectured physical state condition, and one needs to have a more fundamental derivation of this physical state condition in order to make further progress with this approach.
However, this is not an easy task, since it requires the construction of a suitable BRST operator, generalizing the BVW string of \cite{Berkovits:1999im}.
As of this writing, this problem is yet unsolved.
On the other hand, the worldsheet theory for AdS$_3\times$S$^3\times$M$_4$ is on a much more solid footing, and its physical state condition comes from a cohomological argument \cite{Berkovits:1999im}. 
If one could connect the worldsheet theories of AdS$_5$ and AdS$_3$, one might get some useful hints for how to better understand the AdS$_5$ story. 
Indeed, as we will see, the chiral algebra in the \ASF\ case is related to the compactification-independent part of \ASTM, i.e.\ \AST.

\bigskip

In this work, we give a description of the chiral algebra of 4D $\mathcal{N}=4$ $U(N)$ or $SU(N)$  theory from the worldsheet.
In particular, we show the following.
\begin{enumerate}
\item \underline{Spectrum.}
Among the worldsheet free fields that describe 4D $\mathcal{N}=4$ SYM, the Schur condition of the 4D theory selects out \textit{half} of this set; the corresponding worldsheet subsector inherits the same physical state conditions as the full theory and its resulting 4D spectrum contains the Schur subsector of the 4D  theory.
\item \underline{Algebra.} The BPS spectrum of this ``Schur subsector" of the worldsheet theory then generates  an $\mathcal{N}=4$ $\mathcal{W}_{\infty}$ algebra\footnote{
Note that this $\mathcal{N}=4$ $\mathcal{W}_{\infty}$ algebra is not to be confused with the  $\mathcal{N}=4$ $\mathcal{W}_{\infty}$ algebra that appears as the symmetry of the boundary CFT of an $\mathcal{N}=4$ Vasiliev higher spin gravity in $\textrm{AdS}_3$.} that reproduces the chiral algebra computed either directly as in  \cite{Beem:2013sza} or holographically as in \cite{Costello:2018zrm}.
\end{enumerate} 
We emphasize that the free field modes in the Schur subsector actually produce a much bigger algebra than the chiral algebra; this bigger algebra is generated by both short and long multiplets of the 2D $\mathcal{N}=4$ superconformal algebra.
But since we are interested in the chiral algebra, which is independent of the coupling, we should extract the subalgebra that survives even after we turn on the coupling.
One would expect that only the short multiplets (of the 2D $\mathcal{N}=4$ superconformal algebra) within the Schur subsector of the worldsheet CFT are not lifted once the coupling is switched on, and they generate the chiral algebra.
In summary, on the worldsheet the chiral algebra arises simply as the algebra generated by the BPS part of half of the free fields used in the free-field realization of the worldsheet theory.

\medskip

Furthermore, we will show that  the Schur subsector of the worldsheet theory of \ASF\ describes  the ``compactification-independent" part of the worldsheet theory for \ASTM, which can be viewed as an \AST\ $\subset$ \ASF.
This establishes a link between the worldsheet theory for \ASTM\ and the worldsheet theory for \ASF.

\medskip

Finally, the method of this paper can also be applied to those 4D $\mathcal{N}=2$ superconformal theories that are obtained by orbifolding 4D $\mathcal{N}=4$ SYM, whose worldsheet CFTs are also known \cite{Gaberdiel:2022iot}.

\medskip

The plan of this paper is as follows. 
In section 2, we review all the necessary ingredients for this paper.
In section 3, we extract the subsector of the worldsheet CFT of \ASF\ that is selected by the Schur condition and whose corresponding 4D spectrum contains the Schur subsector of 4D $\mathcal{N}=4$ SYM.
In section 4, we first obtain the algebra generated by the physical fields in the worldsheet Schur subsector, and then by restricting to the short multiplets of 2D $\mathcal{N}=4$ superconformal algebra we reproduce the chiral algebra.
In Section 5, we end with a summary and discussion. 

\section{Review}

\subsection{4D \texorpdfstring{$\mathcal{N}=4$}{N=4} SYM, Schur operators, and chiral algebra}

\subsubsection{\texorpdfstring{Large-$N$}{Large-N} spectrum and index}

For the purpose of comparing to the worldsheet theory, we only need to consider the single-particle spectrum (consisting of only single-trace operators). 
The single-trace operators in 4D $\mathcal{N}=4$ SYM are composed of fields from the set of (on-shell) letters \cite{Kutasov:2000td,Polyakov:2001af}
\begin{equation}\label{eq:LetterSet}
S= \{ 
\partial^n\phi^i
\,,\, 
\partial^n\Psi_{A,\alpha}
\,,\, 
\partial^n\Psi_A^{\dot{\alpha}}
\,,\, 
\partial^n{\cal F}_{\alpha \beta}
\,,\, 
\partial^n {\cal F}^{\dot{\alpha} \dot{\beta}} \}\,,
\end{equation}
with $i=1,\dots,6$, $A=1,2,3,4$, $\alpha,\beta, \dot{\alpha},\dot{\beta}=1,2$, and $n\in \mathbb{Z}_{\geq 0}$.
The fields in \eqref{eq:LetterSet} form the singleton representation of the global symmetry $\mathfrak{psu}(2,2|4)$ of the theory: 
\begin{equation}\label{eq:R0}
\begin{aligned}
\mathcal{R}_0 =\bigoplus^{\infty}_{n=0}\big(
(\tfrac{n}{2},\tfrac{n}{2};[0,1,0])_{n+1}
&
\oplus
(\tfrac{n+1}{2},\tfrac{n}{2};[1,0,0])_{n+\tiny{\frac{3}{2}}}
\oplus
(\tfrac{n}{2},\tfrac{n+1}{2};[0,0,1])_{n+\frac{3}{2}} \\
&
\oplus
(\tfrac{n+2}{2},\tfrac{n}{2};[0,0,0])_{n+2}
\oplus
(\tfrac{n}{2},\tfrac{n+2}{2};[0,0,0])_{n+2}
)\big)   \,,     
\end{aligned}
\end{equation}
decomposed in terms of representations $(J_1,J_2,[\lambda_1,\lambda_2,\lambda_3])_E$ of the bosonic subalgebra $\mathfrak{su}(2)_1\oplus \mathfrak{su}(2)_2 \oplus \mathfrak{su}(4)_{R} \subset \mathfrak{psu}(2,2|4)$, where $J_{1,2}$ are the spin of $\mathfrak{su}(2)_{1,2}$,  $[\lambda_1,\lambda_2,\lambda_3]$ is the Dynkin label\footnote{
Here we use the convention that $\mathbf{4}_{s}=[1,0,0]$,  $\mathbf{4}_{c}=[0,0,1]$, and $\mathbf{6}=[0,1,0]$.
} of $\mathfrak{su}(4)_{R}$, and $E$ is the eigenvalue of the dilation operator.

\medskip

The index of the single-trace operators $\IFsp(\mathfrak{q})$ receives contributions from the single-trace operators of length-$w$, whose index is denoted by  $\IFsp^{(w)}(\mathfrak{q})$:
\begin{equation}\label{eq:SPIndex4D}
\IFsp(\mathfrak{q})=\sum^{\infty}_{w=1} \IFsp^{(w)}(\mathfrak{q})\,,
\end{equation}
where $\mathfrak{q}$ stands for the collection of fugacities $\mathfrak{q}=\{q_1,q_2,\dots q_m\}$.\footnote{
Note that the lower bound in the sum, $w=1$, is for the $U(N)$ theory. 
If we are considering $SU(N)$, then 
the lower bound in the sum  would be $w=2$.
}
The problem of counting  the single-trace operators with length-$w$ is then a special case of Polya's enumeration theorem \cite{Polya1987} with the finite group being the cyclic group $\mathbb{Z}_n$,\footnote{
This combinatorics problem can be phrased as counting the configurations of a necklace of length-$n$ with beads colored by the states from the set \eqref{eq:LetterSet}.}  which relates  $\IFsp^{(w)}(\mathfrak{q})$ to the index of one letter from the set \eqref{eq:LetterSet}, denoted by  $\IFsl(\mathfrak{q})$ \cite{Polyakov:2001af, Bianchi:2003wx}:
\begin{equation}\label{eq:SPIndex4Dw}
\IFsp^{(w)}(\mathfrak{q})=\frac{1}{w} \sum_{d|w}\, \phi(d)\, \IFsl(\mathfrak{q}^d)^{\frac{w}{d}}\,,
\end{equation}
where Euler's totient function $\phi(d)$ counts the number of order-$d$ elements in the cyclic group $\mathbb{Z}_w$ and can be computed by $\phi(d)=\sum_{k|d} k\,\mu(\frac{d}{k})$ where $\mu(n)$ is the M\"{o}bius function.\footnote{
Recall that $\mu(n)=1$ when $n=1$, $\mu(n)=(-1)^k$ when $n$ is a product of $k$ distinct primes, and $\mu(n)=0$ otherwise.
In particular, $\mu(p)=-1$ for $p$ prime.
}

\medskip

Consider the supercharge $\mathcal{Q}$ with spin
\begin{equation}
(j_1,j_2)\equiv  (J^3_1,J^3_2)=(-\frac{1}{2},0)    \,.
\end{equation} 
The single-letter index defined by  $\mathcal{Q}$ is \cite{Kinney:2005ej}:
\begin{equation}\label{eq:indexN4vec0}
\begin{aligned}
\IFsl_{\textrm{vec}}(T,a_2,v_2,v_3)&=\textrm{Tr}_{\mathcal{R}_0}(-1)^F T^{2(E+j_1)}a_2^{2j_2}v_2^{R_2}v_3^{R_3}\,,
\end{aligned}
\end{equation}
where $R_{1,2,3}$ are the three Cartan generators of $\mathfrak{su}(4)_R$. 
It selects those states in \eqref{eq:R0} that satisfy:
\begin{equation}\label{eq:16BPS}
\Delta\equiv 2\{\mathcal{Q}^{\dagger},\mathcal{Q}\}=E-(2j_1+\frac{3R_1+2R_2+R_3}{2}  )=0  \,.
\end{equation}
The condition \eqref{eq:16BPS} allows one to evaluate the index explicitly, as in \cite{Kinney:2005ej}. 
To compare to the literature on the chiral algebra such as \cite{Beem:2013sza}, we recast the result of \cite{Kinney:2005ej} in the $\mathcal{N}=2$ language:
\begin{equation}\label{eq:indexN4vec}
\begin{aligned}
\IFsl_{\textrm{vec}}(a,p,q,t)&=
\textrm{Tr}_{\mathcal{R}_0}(-1)^Fp^{\tiny{\frac{1}{2}(E-2j_2-2R-r)}}q^{\tiny{\frac{1}{2}(E+2j_2-2R-r)}}a^{R_2/2}t^{R+r}\\
&=1-\frac{(1-a^{\tiny{\frac{1}{2}}} t^{\tiny{\frac{1}{2}}})
(1-a^{-\tiny{\frac{1}{2}}} t^{\tiny{\frac{1}{2}}})
(1-(pq) t^{-1})}{(1-p)(1-q)}\,,
\end{aligned}
\end{equation}
where we have redefined the fugacities
\begin{equation}
T=(pq)^{\tiny{\frac{1}{6}}}
\,, \quad 
a_2=(p^{-1}q)^{\tiny{\frac{1}{2}}}
\,, \quad
v_2=(pq)^{\tiny{-\frac{1}{3}}}a^{\tiny{\frac{1}{2}}}t^{\tiny{\frac{1}{2}}}
\,, \quad
v_3=(pq)^{\tiny{-\frac{2}{3}}}t\,,
\end{equation}
and used the condition \eqref{eq:16BPS}; $R$ and $r$ are the charge of the $\mathcal{N}=2$ R-symmetry $\mathfrak{su}(2)_R$ and $\mathfrak{u}(1)_r$ symmetry, respectively, and are  related to the $\mathfrak{su}(4)_R$ Cartan generators $R_{1,2,3}$ by
\begin{equation}\label{eq:Rrdef}
R=\frac{1}{2}(R_1+R_2+R_3)
\qquad \textrm{and} \qquad
r=\frac{-R_1+R_3}{2}\,.
\end{equation}
Finally $a$ is the fugacity of the $\mathfrak{su}(2)_F$ flavor symmetry, which is the commutant of $\mathfrak{su}(2)_R\oplus\mathfrak{u}(1)_r\subset \mathfrak{su}(4)_R$; its charge is related to the $\mathfrak{su}(4)_R$ Cartans by $R_F=\frac{R_2}{2}$.

\medskip

The full index of $\mathcal{N}=4$ U(N) SYM is
\begin{equation}
\IFf^{U(N)}(a,p,q,t)=\frac{1}{|N!|}\oint [d\vec{b}]\,\Delta(\vec{b})\, \textrm{PE}[\, \IFsl_{\textrm{vec}}(a,p,q,t)\, \chi_{\textrm{adj}}(\vec{b})\, ] \,,
\end{equation}
where
$\chi_{\textrm{adj}}(\vec{b})$ is the character of the adjoint representation of $U(N)$.
In the large-$N$ limit, the integration greatly simplifies since the matrix integral is captured by the zero modes, and the integral is given by the one-loop determinant; it can be evaluated exactly and gives 
\begin{equation}\label{eq:IndexLargeNUN}
N\rightarrow \infty: \qquad    \IFf^{U(N)}(a,p,q,t)=\prod^{\infty}_{k=1}\frac{1}{1-\IFsl_{\textrm{vec}}(a^k,p^k,q^k,t^k)}\,.
\end{equation}

\subsubsection{Schur index and Schur operators
}

The Schur limit is defined as $t\rightarrow q$, under which the single-letter index of the $\mathcal{N}=4$ vector multiplet \eqref{eq:indexN4vec} becomes\footnote{
One can show that the Schur index is independent of $p$ and hence one can set $p\rightarrow 0$.
}
\begin{equation}\label{eq:indexVecN4Schur} 
\begin{aligned}
\IFsl^{\textrm{Schur}}_{\textrm{vec}}(a,q)
= \textrm{lim}_{ t\rightarrow q,p\rightarrow 0}   \IFsl_{\textrm{vec}}(a,p,q,t)
%&=1-\frac{(1-s\sqrt{q} )    (1-s^{-1} \sqrt{q})}{(1-q)}\\
&=\frac{\sqrt{q}}{1-q}\chi_{\frac{1}{2}}(a)-\frac{2 q}{1-q}\,,
\end{aligned}
\end{equation}
 where $\chi_j(a)$ is the character of the spin-$j$ representation of $\mathfrak{su}(2)_F$: $\chi_j(a)=\sum^{j}_{m=-j}a^m$.
The Schur limit selects operators (the so-called Schur operators) that satisfy
\begin{equation}\label{eq:SchurCondition}
E= (j_1+j_2)+2R \qquad \textrm{and} \qquad r=j_1-j_2\,.
\end{equation}
From the single-letter Schur index \eqref{eq:indexVecN4Schur}, one can compute the Schur index of the $\mathcal{N}=4$ $U(N)$ theory in the large-$N$ limit:
\begin{equation}\label{eq:IndexLargeNSchurUN}
\begin{aligned}
N\rightarrow \infty: \quad    \IFf^{\textrm{Schur},U(N)}(a,q)
&=\prod^{\infty}_{k=1}\frac{1}{1-\IFsl^{\textrm{Schur}}_{\textrm{vec}}(a^k,q^k)} \,.
\end{aligned}
\end{equation}
Similarly, the Schur index of the $\mathcal{N}=4$ $SU(N)$ theory in the large-$N$ limit is
\begin{equation}\label{eq:IndexLargeNSchurSUN}
N\rightarrow \infty: \qquad   \IFf^{\textrm{Schur},SU(N)}(a,q)=\prod^{\infty}_{k=1}\frac{\textrm{Exp}[-\frac{1}{k}\IFsl^{\textrm{Schur}}_{\textrm{vec}}(a^k,q^k)]}{1-\IFsl^{\textrm{Schur}}_{\textrm{vec}}(a^k,q^k)}\,.\nonumber
\end{equation}

\subsubsection{Chiral algebra}

It was shown in \cite{Beem:2013sza} that for a general 4D $\mathcal{N}=2$ superconformal theory, by considering the BRST cohomology of $\mathcal{Q}+\mathcal{S}$ where $Q$ is one of the supercharges and $S$ is the superconformal charge, we select the Schur subsector restricted to $\mathbb{R}^2\subset \mathbb{R}^4$.
These Schur local operators are governed by an (infinitely dimensional) vertex operator algebra.
In particular, the character of the chiral algebra reproduces the Schur index of the 4D theory,\footnote{One can view the chiral algebra as the categorification of the Schur index of the 4D theory.} and the chiral algebra determines the correlation functions of the Schur operators, as meromorphic functions of the $\mathbb{R}^2$ coordinates.

\medskip

In this paper, we will not discuss the correlation functions but only focus on the Schur index. 
The full Schur index for the $U(N)$ theory  \eqref{eq:IndexLargeNSchurUN} has the expansion
\begin{equation}\label{eq:IndexLargeNSchurProduct}
\begin{aligned}
&\IFf^{\textrm{Schur},U(N)}(a,q)
=1
+q^{\frac{1}{2}}\,\chi_{\frac{1}{2}}(a)
+q\,(2\chi_{1}(a)-2\chi_{0}(a))+q^{\frac{3}{2}}\,(3\chi_{\frac{3}{2}}(a)-2\chi_{\frac{1}{2}}(a))
\\
&+q^{2}\,(5\chi_{2}(a)-4\chi_{1}(a)+\chi_{0}(a))
+q^{\frac{5}{2}}\,(7\chi_{\frac{5}{2}}(a)-5\chi_{\frac{3}{2}}(a)+\chi_{\frac{1}{2}}(a))+\dots\,,
\end{aligned}
\end{equation}
which should match the full index of the vacuum representation of a certain $\mathcal{N}=4$ $\mathcal{W}_{\infty}$ algebra that is the chiral algebra of the $U(N)$ theory.\footnote{Note that this is not to be confused with the $\mathcal{N}=4$ $\mathcal{W}_{\infty}$ algebra that arises as the boundary symmetry of the high spin gravity in AdS$_3$, see also the discussion in Section.\ \ref{ssec:chiralVShigherspin}.}
Similarly, the expansion of the full Schur index for the $SU(N)$ theory \eqref{eq:IndexLargeNSchurSUN} is
\begin{equation}\label{eq:IndexLargeNSchurExp}
\begin{aligned}
&\IFf^{\textrm{Schur},SU(N)}(a,q)
=1
+q\,\chi_{1}(a)
+q^{\frac{3}{2}}\,(\chi_{\frac{3}{2}}(a)-2\chi_{\frac{1}{2}}(a))
 +q^{2}\,(2\chi_{2}(a)-\chi_{1}(a)+2\chi_{0}(a))\\
&+q^{\frac{5}{2}}\,(2\chi_{\frac{5}{2}}(a)-2\chi_{\frac{3}{2}}(a)-2\chi_{\frac{1}{2}}(a))
 +q^3\,(4\chi_{3}(a)-3\chi_{2}(a)+2\chi_{1}(a)+3\chi_{0}(a))+\dots
\end{aligned}
\end{equation}
which should match the index of the vacuum representation of the chiral algebra of the $SU(N)$ theory.
We will explain how to reproduce them from the worldsheet perspective later.

\subsubsection{Twisted holography}

The gravity dual of 4D $\mathcal{N}=4$ SYM is IIB supergravity on \ASF.
A natural question is how to describe the gravity dual of its chiral algebra.
This was answered by twisted holography \cite{Costello:2018zrm}, which we briefly review below.
For subsequent developments see e.g.\  \cite{Budzik:2021fyh,Budzik:2022hcd,Costello:2023hmi}

In this set up, one starts with the full holographic duality arising from type IIB string in flat space together with a stack of $N$ D3-branes: in the large-$N$ limit, turning on the backreaction gives IIB string theory living on \ASF\ as the gravity dual of 4D $\mathcal{N}=4$ SYM.
Before the backreaction, twisting the whole setup in the presence of an $\Omega$-background localizes to the B-model topological string on $\mathbb{C}^3$, with $N$ D2-branes\footnote{The D2-branes in the B-model correspond to Lorentzian D1-branes and have a two-dimensional worldvolume.} wrapping  $\mathbb{C}$ in $\mathbb{C}^3$.
The B-brane worldvolume theory is the gauged $\beta \gamma$ system $\mathcal{A}_N$ (in adjoint of $U(N)$)  \cite{Witten:1992fb} and gives the chiral algebra of 4D $\mathcal{N}=4$ SYM with $G=U(N)$, see the review in \cite[App.\ E]{Costello:2018zrm}.
Finally, turning on the backreaction in the large-$N$ limit, one obtains the B-model topological string on SL$(2,\mathbb{C})$ \cite{Costello:2018zrm}.
In \cite{Costello:2018zrm,Budzik:2021fyh,Budzik:2022hcd}, the computation was mostly done in the bulk, namely in terms of Kodaira-Spencer theory (which is a topological version of the closed string field theory) rather than in terms of the worldsheet theory.

Note that \cite{Costello:2018zrm} bypassed the actual twisting procedure of \cite{Costello:2016mgj} (namely the counterpart of the twisting procedure of \cite{Beem:2013sza} in the 4D theory) and instead obtained the end-result of the twisting by directly considering the worldvolume theory of D2-branes in the B-model, which gives the chiral algebra \cite{Witten:1992fb}. 
Similarly, in the current paper we also bypass the actual twisting procedure at the level of the worldsheet theory but instead start by extracting the worldsheet subsector that is selected by the Schur condition (which we will call ``worldsheet Schur subsector") and whose corresponding 4D spectrum contains the Schur subsector of the 4D theory and then show that the short multiplets in the worldsheet Schur subsector reproduce the chiral algebra.

\subsection{Worldsheet theory of \texorpdfstring{\ASTT}{AdS3S3T4}}
\label{ssec:AdS3worldsheet}

Although our main focus is on the worldsheet theory of \ASF, we first briefly review the worldsheet theory of \ASTT, since the construction of  the former is largely inspired by the latter, which is also much further developed as of this writing.

\subsubsection{Free field relation of \texorpdfstring{$\mathfrak{psu}(1,1|2)_1$}{psu(1,1|2)1}}

The worldsheet CFT of  \ASTT\ that is dual to the (free) symmetric orbifold of $T^4$ 
was proposed in \cite{Eberhardt:2018ouy}.
The main ingredient is a free field realization of the current algebra $\mathfrak{psu}(1,1|2)_1$.
The reason that it is level-$1$ is because the worldsheet dual of the symmetric orbifold of $T^4$ should correspond to a string theory at the tensionless limit. 
Hence the AdS$_3$ radius should be as small as possible in string units, which means that the level of the current algebra should take the smallest possible value, which is one in this case.  
The set of free fields  consists of\footnote{
We largely follow the notation of \cite{Eberhardt:2018ouy} for the \ASTT\ case and that of \cite{Gaberdiel:2021qbb} for the \ASF\ case, except that one of the complex fermions in the  \ASTT\ case that is labeled as $\psi$ in \cite{Eberhardt:2018ouy} is here relabeled as $\phi$ to avoid conflict with the \ASF\ case.} 
\begin{equation}\label{eq:FreeFieldAdS3}
4\textrm{ symplectic bosons } (\xi^\pm, \eta^\pm)
\qquad\textrm{and}\qquad 
2 \textrm{ complex fermions } (\phi^\pm, \chi^\pm)\,.
\end{equation}
They all have conformal weight $h=\frac{1}{2}$, and satisfy the mode relations\footnote{
$\epsilon$ is the totally symmetric tensor and the convention here is that $\epsilon^{+-} = 1$.}
\begin{equation}\label{eq:free_field_CR}
[\xi^\alpha_r,\eta^\beta_s] = \epsilon^{\alpha\beta}\delta_{r+s,0}
\,,\quad \{\phi^\alpha_r,\chi^\beta_s\} = \epsilon^{\alpha\beta}\delta_{r+s,0}
\,.
\end{equation}
The neutral bilinears of these fields generate the current algebra $\mathfrak{u}(1,1|2)_1$, for more details see \cite[App.\ C]{Eberhardt:2018ouy}.
To describe the worldsheet theory of \ASTT, we really need the current algebra $\mathfrak{psu}(1,1|2)_1$, which is obtained from  $\mathfrak{u}(1,1|2)_1$ upon setting $Z_m$ in \eqref{eq:ZY} to zero, see \cite{Dei:2020zui} for more details.

The bosonic subalgebra of  $\mathfrak{psu}(1,1|2)_1$ is
\begin{equation}
\mathfrak{sl}(2,\mathbb{R})_{1}\oplus \mathfrak{su}(2)_{1}\,.
\end{equation}
The first factor $\mathfrak{sl}(2,\mathbb{R})_{1}$ is realized by\footnote{
Recall that $\mathfrak{sl}(2,\mathbb{R})_{k}=\mathfrak{su}(2)_{-k}$, and $\mathfrak{su}(2)_{k}$ is defined as \begin{equation}
{}[J^3_m,J^3_n]  = \frac{k}{2}  m \delta_{m+n,0} 
\,, \quad 
{}[J^3_m,J^\pm_n]  =  \pm J^\pm_{m+n} 
\,, \quad 
{}[J^+_m,J^-_n]  = 2  J^3_{m+n} + k  m \delta_{m+n,0}\,.\nonumber
\end{equation}}
\begin{equation}
J^3_m=-\tfrac{1}{2} (\eta^+\xi^{-}+\eta^-\xi^+)_m
\qquad \textrm{and} \qquad 
J^\pm_m=(\eta^\pm\xi^\pm)_m\,.
\end{equation}
Similarly, the second factor $\mathfrak{su}(2)_{1}$  is realized as
\begin{equation}
K^3_m=-\tfrac{1}{2} (\chi^+\psi^{-}+\chi^-\psi^+)_m
 \qquad \textrm{and} \qquad 
K^\pm_m=\pm(\chi^\pm\psi^\pm)_m\,.
\end{equation}
The fermionic generators of $\mathfrak{psu}(1,1|2)_1$ are 
\begin{equation}
S_m^{\alpha\beta+}=(\chi^\beta \xi^\alpha)_m
 \qquad \textrm{and} \qquad 
 S_m^{\alpha\beta-}=-(\eta^\alpha\psi^\beta)_m \,.
\end{equation}
In addition, we define
\begin{equation}
U_m=-\tfrac{1}{2} (\eta^+\xi^- - \eta^-\xi^+)_m
\qquad \textrm{and}
\qquad
 V_m=-\tfrac{1}{2} (\chi^+\psi^- - \chi^-\psi^+)_m\,,
\end{equation}
and also 
\begin{equation}\label{eq:ZY}
Z_m=U_m+V_m 
\qquad \textrm{and} \qquad 
Y_m=U_m-V_m\,.
\end{equation}

\subsubsection{Spectral flow and physical states}
\label{sssec:AdS3phycicalcondition}

To match the field theory spectrum, one needs to impose the physical state conditions on the worldsheet fields, which was derived from the BRST cohomology of the hybrid formalism of Berkovits-Vafa-Witten,  see \cite{Eberhardt:2018ouy} for this derivation and the full list of physical state conditions. 
After the physical state conditions are imposed,  the partition function of the resulting physical spectrum can be written as a sum over all the spectrally-flowed sectors, labeled by $w$:
\begin{equation}
\pfWS^{\textrm{w.s.}}_{\textrm{\ASTT}}(\mathfrak{q})=\sum^{\infty}_{w=1} \pfWS^{(w)}_{\textrm{\ASTT}}(\mathfrak{q})    \,,
\end{equation}
where  the partition function $\pfWS^{(w)}(\mathfrak{q}) $ from the $w$-spectrally-flowed sector reproduces the single-particle spectrum of the $w$-cycle twisted sector of the symmetric orbifold of $T^4$ \cite{Eberhardt:2018ouy}.

\subsection{Worldsheet theory of \texorpdfstring{AdS$_5\times S^5$}{AdS5S5}}

\subsubsection{Free field relation of \texorpdfstring{$\mathfrak{psu}(2,2|4)_1$}{ps2{2,2|4}1}}

In view of the fact that the $N=4$ SYM has the superconformal symmetry $\mathfrak{psu}(2,2|4)$, the starting point of the proposal in \cite{Gaberdiel:2021qbb, Gaberdiel:2021jrv} is the free field realization of the current algebra $\mathfrak{psu}(2,2|4)_1$,\footnote{The reason why the level is one is same as for \ASTT, see Section.\ \ref{sssec:AdS3phycicalcondition}.} with the set of free fields consisting of $8$ symplectic bosons
\begin{equation}\label{eq:8SB} 
(\lambda^\alpha\,,
\lambda^{\dagger}_{\dot{\alpha}}\,,
 \mu^{\dot{\alpha}}\,, 
\mu^{\dagger}_{\alpha})
\qquad \textrm{with}\quad 
 \alpha,\dot{\alpha}=1,2\,,
\end{equation} 
and four complex fermions 
\begin{equation}\label{eq:4CF}
(\psi^a\,,\psi^\dagger_a) 
\qquad \textrm{with} \quad 
a=1,2,3,4\,.
\end{equation}
It will be convenient to group the free fields according to their indices:
\begin{equation}\label{eq:YZlist}
Y_I=(\mu^{\dagger}_{\alpha}, \lambda^{\dagger}_{\dot{\alpha}},\psi^{\dagger}_{a})
\qquad \textrm{and} \qquad 
Z^I=(\mu^{\dot{\alpha}}, \lambda^{\alpha},\psi^{a})\,.
\end{equation}
All the fields have conformal dimension $h=\frac{1}{2}$.
The commutation relations among their modes are
\begin{equation}\label{ffcomm}
[\lambda^\alpha_r,(\mu^\dagger_\beta)_s] = \delta^{\alpha}_{\beta} \, \delta_{r+s,0} 
\,, \qquad 
[\mu^{\dot{\alpha}}_r,(\lambda^\dagger_{\dot{\beta}})_s] = \delta^{\dot{\alpha}}_{\dot{\beta}} \, \delta_{r+s,0}
 \,, \qquad 
\{\psi^a_r,(\psi^\dagger_b)_s \} = \delta^{a}_b \, \delta_{r+s,0} \,,
\end{equation}
from which one can show that they generate the current algebra $\mathfrak{psu}(2,2|4)_1$ \cite{Gaberdiel:2021qbb, Gaberdiel:2021jrv}, as we will review below.

\medskip

First of all, the compact bosonic subalgebra of $\mathfrak{psu}(2,2|4)_1$:
\begin{equation} \mathfrak{su}(2)_{-1} \oplus \mathfrak{su}(2)_{-1}\oplus \mathfrak{su}(4)_{1} \subset \mathfrak{psu}(2,2|4)_1
\end{equation}
is realized as follows.
The first $\mathfrak{su}(2)_{-1}$ factor  is generated by
\begin{equation}\label{eq:Jsu2}
J^3_m = \frac{(\mu^\dagger_2 \, \lambda^2)_m- (\mu^\dagger_1 \, \lambda^1)_m}{2} 
\,, \qquad
J^+_m = (\mu^\dagger_2 \, \lambda^1)_m \ , \qquad J^-_m =   (\mu^\dagger_1 \, \lambda^2)_m \,,
\end{equation}
where all the products are normal ordered.
Similarly, the second $\mathfrak{su}(2)_{-1}$ is generated by 
\begin{equation}\label{eq:Jdotsu2}
\dot{J}^3_m = \frac{(\lambda^\dagger_2 \, \mu^2)_m- (\lambda^\dagger_1 \, \mu^1)_m}{2} 
\,, \qquad
\dot{J}^+_m = (\lambda^\dagger_2 \, \mu^1)_m \ , \qquad 
\dot{J}^-_m =   (\lambda^\dagger_1 \, \mu^2)_m \,.
\end{equation}
Finally, the factor $\mathfrak{su}(4)_{1}$ is generated by
\begin{equation}
({\cal R}^a{}_b)_m =( \psi^\dagger_b \psi^a)_m- \tfrac{1}{4} \delta^a_b (\psi^\dagger_c \psi^c )_m \,,
\end{equation}
where we use the convention in which the positive roots of $\mathfrak{su}(4)$ are given by  $({\cal R}^{a}{}_b)_0$ with $a<b$.
In particular,  the three Cartan generators of $\mathfrak{su}(4)$ are 
\begin{equation}\label{eq:cartansu4}
H_1  =  ( \psi^\dagger_2 \psi^2)_0  -  ( \psi^\dagger_1 \psi^1)_0
\ , \quad 
H_2  =  ( \psi^\dagger_3 \psi^3)_0  -  ( \psi^\dagger_2 \psi^2)_0
\ , \quad
H_3  =  ( \psi^\dagger_4 \psi^4)_0  -  ( \psi^\dagger_3 \psi^3)_0 \ .
\end{equation}
The non-compact bosonic generators of $\mathfrak{psu}(2,2|4)_1$ are
\begin{equation}
 \mathcal{P}^{\dot{\alpha}}{}_{\beta}  =  \mu^{\dot{\alpha}}\, \mu^\dagger_\beta 
\qquad \textrm{and} \qquad 
\mathcal{K}^{\alpha}{}_{\dot{\beta}}  =  \lambda^\alpha \, \lambda^\dagger_{\dot{\beta}} \,,
\end{equation}
which are the translation and  special conformal generators, respectively. 
Finally, the fermionic generators are
\begin{equation}
\begin{aligned}
& {\cal S}^\alpha{}_a  =  \lambda^\alpha\, \psi^\dagger_a
\,, \quad 
 \dot{\cal S}^a{}_{\dot{\alpha}}  =   \psi^a \, \lambda^\dagger_{\dot{\alpha}}  
\qquad \textrm{and}
\qquad
\dot{\cal Q}^{\dot{\alpha}}{}_{a}  =  \mu^{\dot{\alpha}} \, \psi^\dagger_a
\,, \quad 
{\cal Q}^a{}_\alpha =  \psi^a\, \mu^\dagger_\alpha \,.
\end{aligned}
\end{equation}

\medskip

However, the list of fields
\begin{equation}\label{eq:psulist}
J\,, \dot{J}\,, \mathcal{R}\,, \mathcal{K}\,, \mathcal{P}\,, \mathcal{S}\,, \dot{\mathcal{S}}\,, \mathcal{Q}\,, \dot{\mathcal{Q}}
\end{equation}
 defined above do not close upon themselves and hence do not form the $\mathfrak{psu}(2,2|4)_1$ algebra at face value. 
Instead, the $\mathfrak{psu}(2,2|4)_1$ algebra  can be obtained from the larger $\mathfrak{u}(2,2|4)_1$ as follows.
Apart from the fields in the list \eqref{eq:psulist}, $\mathfrak{u}(2,2|4)_1$ has two additional (bosonic) generators:
\begin{equation}
\mathcal{B}_m  =   \tfrac{1}{2} ( \mu^\dagger_\alpha \, \lambda^\alpha + \lambda^\dagger_{\dot{\alpha} }\mu^{\dot{\alpha}})_m
\qquad \textrm{and}\qquad 
\mathcal{C}_m
=  \tfrac{1}{2} ( \mu^\dagger_\alpha \, \lambda^\alpha + \lambda^\dagger_{\dot{\alpha} }\mu^{\dot{\alpha}} +\psi^{\dagger}_a\psi^a)_m  \,. 
\end{equation}
From  their commutation relations (see \cite[App.\ A]{Gaberdiel:2021jrv}), one can show that the (anti-)commutation relations of the fields in \eqref{eq:psulist} do not close upon themselves, but also contain $\mathcal{C}$ (but not $\mathcal{B}$).
In addition, we have
\begin{equation}
[\mathcal{C},\textrm{fields in \eqref{eq:psulist}}]=0
\,, \quad 
[\mathcal{C},\mathcal{B}]=\textrm{central}
\,, \quad
[\mathcal{B}, \textrm{fermions in \eqref{eq:psulist}}]\neq 0\,.
\end{equation}
Therefore, one can first obtain an $\mathfrak{su}(2,2|4)_1$ algebra generated by the fields in \eqref{eq:psulist} and $\mathcal{C}$, in which $\mathcal{C}$ is central; then to obtain the $\mathfrak{psu}(2,2|4)_1$ algebra, one simply takes all the fields in \eqref{eq:psulist} and mods out $\mathcal{C}$ by imposing the condition\footnote{
More precisely, at the level of the states, this can be done by  imposing $
{\cal C}_n =0$ with $
n\geq 0$, and the ${\cal C}_{-n}$ descendants are then null and are naturally quotiented out. (This follows a similar argument as for the AdS$_3$ case in \cite{Dei:2020zui}.)}
\begin{equation}
\mathcal{C}=0\,,
\end{equation}
which is the so-called anbitwistor constraint.
For later convenience, we note that 
\begin{equation}
\mathcal{C}=\frac{1}{2}Y_I Z^I\,,
\end{equation}
from which it is transparent that the $Y_I$'s have $\mathcal{C}$-charge $\frac{1}{2}$ whereas the $Z^{I}$'s have $\mathcal{C}$-charge $-\tfrac{1}{2}$, and all the generators of $\mathfrak{psu}(2,2|4)_1$ are $\mathcal{C}$-charge neutral.

\medskip

From these commutation relations, one can deduce that 
\begin{equation}
{\cal D}_0  =  \tfrac{1}{2} \, ( \mu^\dagger_\alpha \, \lambda^\alpha - \lambda^\dagger_{\dot{\alpha} }\, \mu^{\dot{\alpha} })_0
\end{equation}
serves as the dilatation operator of  4D $\mathcal{N}=4$ SYM.

\medskip

Finally, note that we just described the left-movers, and there is another copy for the right-moving sector.
The physical state condition postulated in \cite{Gaberdiel:2021jrv,Gaberdiel:2021qbb} ensures that only the left-movers survive to contribute to the physical spectrum. 
In the end, the physical spectrum of the worldsheet theory consists of Ramond sector together with its spectrally-flowed sectors and matches the spectrum of 4D $\mathcal{N}=4$ SYM.

\subsubsection{Spectral flow and physical states}
\label{ssec:SpectralFlowAdS5}

Summarizing the proposal of \cite{Gaberdiel:2021qbb}, the procedure for obtaining the physical spectrum is as follows.
\begin{enumerate}
\item In the $w$-spectrally-flowed sector, consider the subset of 
the free fields \eqref{eq:YZlist}:
\begin{equation}\label{eq:WedgeSubset}
\begin{aligned}
Y_I \,\, \supset \,\,
{}^{\vee}Y_I
&=(
\mu^{\dagger}_1
\,, \, 
\mu^{\dagger}_2
\,, \, 
\psi^{\dagger}_1 
\,,\, 
\psi^{\dagger}_2) 
\,,  \\
Z^I \,\, \supset \,\, 
{}^{\vee}Z^I
&=(\mu^1
\,, \,
\mu^2
\,, \,
\psi^3 
\,,\,
\psi^4) \,,
\end{aligned}
\end{equation}
and restrict to the space generated by the ``wedge modes" of this subset:\footnote{
They are called wedge modes since as $w$ runs over $\mathbb{N}$, the condition $-\tfrac{w-1}{2}\le r \le \tfrac{w-1}{2}$ takes the shape of a wedge. 
This is not to be confused with the wedge modes of $\mathcal{W}$ algebras although these two nomenclatures share the same origin.  
} 
\begin{equation}\label{eq:WedgeCondition}
(\mu^\dagger_{1,2})_r 
\ , \quad 
(\mu^{1,2})_r 
\ , \quad   
(\psi^\dagger_{1,2})_r  
\ ,  
\quad (\psi^{3,4})_r 
\qquad \textrm{with} \quad 
-\tfrac{w-1}{2} \leq r \leq \tfrac{w-1}{2}\,,
\end{equation}
acting on the ground state $|0\rangle_{w}$.
(Note that only the left-movers of the worldsheet CFT are included.)
\item On this space, impose the residual Virasoro constraint 
\begin{equation}\label{eq:L0constraint}
(L_0+n w)\, |\Psi_{\rm phy}\rangle =0  
\quad \textrm{with} \quad
n\in \mathbb{Z} \,,
\end{equation}
where
\begin{equation}
[L_0\,,\, (\mu^\dagger_{1,2})_r] = - r (\mu^\dagger_{1,2})_r  \,, 
\end{equation}
and similarly for all the other wedge modes, and 
\begin{equation}
L_0 |0\rangle_w = \frac{w}{2} |0\rangle_w\,.
\end{equation}
This condition corresponds to the (spacetime) momentum conservation up to cyclicity. 
\item Impose the``central term" constraint
\begin{equation}\label{eq:Cconstraint}
{\cal C}_n \, |\Psi_{\rm phy}\rangle =0 
\quad \textrm{with} \quad
n=0,1,\ldots , w-1 \,. 
\end{equation}
\end{enumerate}
As was observed in \cite[Section.\ 4.2]{Gaberdiel:2021jrv}, the wedge constraint and the $\mathcal{C}_n=0$ constraint together restrict us to states that are generated from the ground state by the $\mathcal{C}$-neutral DDF-like operators:
\begin{equation}\label{eq:DDF}
(S_I^{\,\,J})_m=\sum^{\frac{w-1}{2}}_{r=m-\frac{w-1}{2}}({}^{\vee}Y_I)_r({}^{\vee}Z^J)_{m-r}
\quad \textrm{with} \quad 
m=0,1,2,\dots, w-1 \,,
\end{equation}
in the $w$-spectrally-flowed sector, where ${}^{\vee}Y_I$ and ${}^{\vee}Z^J$ are from the list \eqref{eq:WedgeSubset} and their modes numbers are inside the wedge \eqref{eq:WedgeCondition}.
One can check that 
\begin{equation}
[L_0\,, \,  (S_I^{\,\,J})_m]=-m \,  (S_I^{\,\,J})_m \,.
\end{equation}
Therefore, the $L_0$ constraint \eqref{eq:L0constraint} dictates that in the  $w$-spectrally-flowed sector, the physical states are generated by products of DDF operators
with total zero momentum up to cyclicity:
\begin{equation}
\prod_i (S_I^{\,\,J})_{m_i} 
\quad \textrm{with}\quad
\sum_i m_i=0 \quad \textrm{mod}\,\, w\,,
\end{equation}
acting on the ground state $|0\rangle_w$.

\bigskip

After imposing the physical state condition, the spectrum of the worldsheet theory for \ASF\ is also given by a sum over all $w$-spectrally-flowed sectors:
\begin{equation}
\pfWS^{\textrm{w.s.}}_{\textrm{\ASF}}(\mathfrak{q})=\sum^{\infty}_{w=1} \pfWS^{(w)}_{\textrm{\ASF}}(\mathfrak{q})    \,,
\end{equation}
where $\mathfrak{q}$ stands for all the fugacities   collectively.
For $w=1$,
\begin{equation}
\pfWS^{(1)}(\mathfrak{q}) =\textrm{Tr}_{\mathcal{H}}[\mathfrak{q}^{\mathfrak{Q}}]=:\pfWS(\mathfrak{q})
\end{equation}
is the character of the RR vacuum with $\mathfrak{Q}$ denoting the charges collectively. 
We also define
\begin{equation}
\widetilde{\pfWS}(\mathfrak{q})=\textrm{Tr}_{\mathcal{H}}[(-1)^F\mathfrak{q}^{\mathfrak{Q}}]\,.
\end{equation}
The $w$-spectrally-flowed sector $\pfWS^{(w)}(\mathfrak{q})$  captures the cyclically invariant physical states in the $w^{\textrm{th}}$ tensor power of $\pfWS(\mathfrak{q})$:
\begin{equation}\label{eq:ZwCyc}
 \pfWS^{(w)}_{\textrm{\ASF}}(\mathfrak{q})    =\textrm{Tr}_{\mathcal{H}^{\otimes w}/\mathbb{Z}_w}[\mathfrak{q}^{\mathfrak{Q}}]
=\frac{1}{w}\sum^{w-1}_{k=0}\pfWS^{\sigma^{k}}(\mathfrak{q}) 
=\frac{1}{w}\sum^{w-1}_{k=0}\textrm{Tr}_{\mathcal{H}^{\otimes w}}[\mathfrak{q}^{\mathfrak{Q}}\sigma^k]\,,
\end{equation} 
where $\sigma=(12\dots w)$ is the cyclic permutation of length $w$.
$\pfWS^{(w)}(\mathfrak{q})$ reproduces the single-trace states with $w$ letters in 4D $\mathcal{N}=4$ SYM; here and henceforth we drop the subscript ``\ASF".

The first few $\pfWS^{(w)}(\mathfrak{q})$ are explicitly
\begin{equation}\label{eq:Zw234}
\begin{aligned}
\pfWS^{(2)}(\mathfrak{q})&=\frac{1}{2}\left(
\pfWS(\mathfrak{q})^2
+\IWS(\mathfrak{q}^2)
\right)\,,\\
\pfWS^{(3)}(\mathfrak{q})&=\frac{1}{3}\left(
\pfWS(\mathfrak{q})^3
+2\,\pfWS(\mathfrak{q}^3)
\right)\,,\\
\pfWS^{(4)}(\mathfrak{q})&=\frac{1}{4}\left(
\pfWS(\mathfrak{q})^4
+\IWS(\mathfrak{q}^2)^2
+2\,\IWS(\mathfrak{q}^4
)\right)\,.
\end{aligned}
\end{equation}
For $w\geq 3$ prime, $\pfWS^{(w)}$ has the simple expression of
\begin{equation}\label{eq:Zwprime}
\pfWS^{(w)}(\mathfrak{q})=\frac{1}{w}\left(\pfWS(\mathfrak{q})^w+(w-1)\pfWS(\mathfrak{q}^{w})\right) \qquad \textrm{for } w\geq 3 \textrm{ prime}\,.
\end{equation}

\section{Schur subsector from worldsheet}

In this section, we first explain the matching between the single-particle Schur index of the 4D theory and the index of the untwisted sector of the worldsheet theory. 
We then define a ``Schur subsector" of  the worldsheet theory and show that the physical states from this subsector reproduce the Schur index of the 4D theory. 

\subsection{4D single-particle spectrum v.s.  worldsheet spectrum}

The 4D index \eqref{eq:IndexLargeNUN} or \eqref{eq:IndexLargeNSchurUN} captures the multi-particle spectrum, whereas the worldsheet theory captures only the single-particle part. 
Therefore, in order to compare with the worldsheet result, we need the single-particle spectrum of 4D $\mathcal{N}=4$ SYM \cite{Sundborg:1999ue, Polyakov:2001af, Bianchi:2003wx, Aharony:2003sx}:
\begin{equation}\label{eq:SPIndex4Dsec3}
\IFsp(\mathfrak{q})=\sum^{\infty}_{w=1 \textrm{ or }2} \IFsp^{(w)}(\mathfrak{q})
\qquad \textrm{with} \qquad 
\IFsp^{(w)}(\mathfrak{q})=\frac{1}{w} \sum_{d|w}\, \phi(d)\, \IFsl(\mathfrak{q}^d)^{\frac{w}{d}}\,,
\end{equation}
where the lower bound is $w=1$ for the $U(N)$ theory and $w=2$ for the $SU(N)$ theory.
To confirm that the signs in the index \eqref{eq:SPIndex4Dsec3} are taken care of properly,\footnote{As we will see momentarily, the formulae for the corresponding characters are slightly different.} one can check that the single-particle index \eqref{eq:SPIndex4Dsec3}  is related to the full  index at large-$N$
\begin{equation}\label{eq:SchurLargeN}
\IFf^{U(N)}(\myq)=\prod^{\infty}_{k=1}\frac{1}{1-\IFsl(\myq^k)} 
\qquad \textrm{or} \qquad 
 \IFf^{SU(N)}(a,q)=\prod^{\infty}_{k=1}\frac{\textrm{Exp}[-\frac{1}{k}\IFsl(\mathfrak{q}^k)]}{1-\IFsl(\mathfrak{q}^k)}
\end{equation}
 by
 \begin{equation}
\mathcal{I}_{\textrm{full}}(\mathfrak{q})=\PE[\mathrm{I}_{\textrm{s.p}}(\myq)]
\qquad \textrm{and} \qquad \mathsf{I}_{\textrm{s.p.}}(\myq)=\Plog[ \mathcal{I}_{\textrm{full}}(\mathfrak{q})]\,,
\end{equation}
where  $\PE$ stands for the plethystic exponent and $\Plog$ its inverse, the plethystic log; they are defined as 
\begin{equation}
\PE[f(x)]\equiv \sum^{\infty}_{k=1}\frac{f(x^k)}{k} 
\qquad \textrm{and} \qquad
\Plog[g(x)]\equiv \sum^{\infty}_{n=1}\frac{\mu(n)}{n} \log{g(x^n)}\,.
\end{equation}

Let us now consider the case when the single-letter index in the formulae above is the single-letter Schur index of the $\mathcal{N}=4$ vector-multiplet \eqref{eq:indexVecN4Schur}:
\begin{equation}\label{eq:slIndex4Dsec3}
\IFsl(\myq)=\IFsl^{\textrm{Schur}}_{\textrm{vec}}(a,q)=\frac{\sqrt{q}}{1-q}\chi_{\frac{1}{2}}(a)-\frac{2 q}{1-q}\,.
\end{equation}
The goal of this section is to reproduce the spacetime single-particle Schur index \eqref{eq:SPIndex4Dsec3} with \eqref{eq:slIndex4Dsec3} from the worldsheet theory for \ASF.

\subsection{Schur subsector of worldsheet theory}

Now we directly extract the subsector of the worldsheet theory for \ASF\ that corresponds to the Schur sector. 
Recall that the physical spectrum is generated by the $\mathcal{C}$-neutral DDF-like operators that are bilinears of the free  fields, see \eqref{eq:DDF}.
We would like to impose the Schur condition on these bilinears.
It turns out that they are generated by a subset of the free fields.

\subsubsection{Imposing Schur condition}

In order to impose the Schur condition \eqref{eq:SchurCondition} on the worldsheet spectrum, we first summarize the operators whose eigenvalues appear in \eqref{eq:SchurCondition}, in terms of the worldsheet fields:
\begin{equation}\label{eq:OperatorSet}
\begin{aligned}
\textrm{Dilatation}: \qquad &
\mathcal{D} 
= \tfrac{1}{2} (   \mu^\dagger_\alpha \, \lambda^\alpha
-\lambda^\dagger_{\dot{\alpha}} \, \mu^{\dot{\alpha}}  )_0   \\
\textrm{Cartan of } \mathfrak{su}(2)_1\oplus \mathfrak{su}(2)_1: \qquad &
\begin{cases}
\begin{aligned}
J^3_0 &
=\tfrac{1}{2} (\mu^\dagger_2 \, \lambda^2 - \mu^\dagger_1 \, \lambda^1)_0 \\
\dot{J}^3_0 &
=\tfrac{1}{2}  (\lambda^\dagger_{2} \, \mu^{2} - \lambda^\dagger_{1} \, \mu^{1} )_0 \\
\end{aligned}
\end{cases}\\
\textrm{Cartan of } \mathfrak{su}(4): \qquad &
\begin{cases}
\begin{aligned}
H_1  &
= (\psi^\dagger_2 \psi^2-\psi^\dagger_1 \psi^1)_0 \\
H_2  &
= (\psi^\dagger_3 \psi^3 -  \psi^\dagger_2 \psi^2)_0 \\
H_3  &
= (\psi^\dagger_4 \psi^4-  \psi^\dagger_3 \psi^3)_0  
\end{aligned}
\end{cases}
\end{aligned}
\end{equation}
whose charges are denoted as:
\begin{equation}
\begin{array}{c|c|ccccc|c|c}
\textrm{operator}\,\,\,&\,\,\mathcal{D} \,\,&\,\, (J^{3})_0 \,\,&\,\, (\dot{J}^{3})_0 \,\,& \,\, H_1\,\, &\,\, H_2 \,\,& \,\, H_3 \,\, & \,\, R\,\, & \,\, r\,\,\\
\hline
\textrm{charge}
&E & \left( \right. j_1 \,, & j_2 \,, & \left[\right.R_1 \,,& R_2\,, & R_3 \left.\right]\left. \right) & R& r
\end{array}
\end{equation}
where the charges of the ($\mathcal{N}=2$) $\mathfrak{su}_{R}\oplus \mathfrak{u}(1)_r$ symmetries $R$ and $r$ are defined in \eqref{eq:Rrdef}.

\medskip

Next, we compute the charges of the basic free fields, i.e.\ the 8 symplectic bosons and the 4 complex fermions in \eqref{eq:YZlist}, under the operators in \eqref{eq:OperatorSet}.
In addition, we will also need their charges $\mathcal{C}$
w.r.t.\ the operator $\mathcal{C}_0$.
We summarize the results in the table below
\begin{equation}\label{tabel:FreeFieldCharge}
\begin{array}{|c|c|c|c|c|c|c|c|c|c|}
\hline
&\,\, \,\,
\mathcal{C}
\,\,\,\,&\,\,\,\,
E 
\,\,\,\,&\,\,\,\,
j_1
\,\,\,\,&\,\,\,\,
j_2 
\,\,\,\,&\,\,\,\,
R_1
\,\,\,\,&\,\,\,\,
R_2
\,\,\,\,&\,\,\,\,
R_3
\,\,\,\,&\,\,\,\,
R 
\,\,\,\,&\,\,\,\,
r\,\,\,\,
\\ \hline
\,\,\,\,\boldsymbol{\mu^{\dagger}_{1}} \,\,\,\, &\tfrac{1}{2}  & \tfrac{1}{2} & -\tfrac{1}{2} & 0& 0& 0& 0&0&0 \\ \hline 
\boldsymbol{\textcolor{red}{\mu^{\dagger}_{2}}} & \tfrac{1}{2} &  \tfrac{1}{2}& \tfrac{1}{2} & 0&0 &0 &0 &0& 0\\ \hline 
\textcolor{red}{\lambda^{\dagger}_{1} } &\tfrac{1}{2}  &-\tfrac{1}{2} &0 &-\tfrac{1}{2} &0 &0 &0 &0& 0\\ \hline 
\lambda^{\dagger}_{2} &\tfrac{1}{2}  & -\tfrac{1}{2}&0 &\tfrac{1}{2} & 0&0 &0 &0&0 \\ \hline 
\boldsymbol{\psi^\dagger_1} &\tfrac{1}{2}  & 0&0 &0 & -1&0 &0 &-\tfrac{1}{2}&\tfrac{1}{2} \\ \hline 
\boldsymbol{\textcolor{red}{\psi^\dagger_2} }&\tfrac{1}{2}  &0 & 0& 0&1 & -1& 0&0&-\tfrac{1}{2}  \\ \hline
\textcolor{red}{\psi^\dagger_3} &\tfrac{1}{2} & 0& 0& 0&0 &1 &-1 &0&-\tfrac{1}{2}  \\ \hline 
\psi^\dagger_4 &\tfrac{1}{2} & 0& 0& 0& 0&0 &1 &\tfrac{1}{2}&\tfrac{1}{2}  \\ \hline 
\lambda^1 & -\tfrac{1}{2}  & -\tfrac{1}{2}& \tfrac{1}{2} &0 &0 &0 &0 &0&0 \\ \hline
\textcolor{red}{\lambda^2 } & -\tfrac{1}{2} & -\tfrac{1}{2}&-\tfrac{1}{2} & 0&0 &0 &0 &0& 0\\ \hline
\boldsymbol{\textcolor{red}{\mu^{1}} } &-\tfrac{1}{2}  & \tfrac{1}{2}& 0& \tfrac{1}{2} & 0&0 &0 &0& 0\\ \hline
\boldsymbol{\mu^{2}}  & -\tfrac{1}{2} &  \tfrac{1}{2}&0 & -\tfrac{1}{2} &0 &0 &0 &0&0 \\ \hline
\psi^1 &-\tfrac{1}{2}  &0 &0 & 0& 1&0 &0 &\tfrac{1}{2}& -\tfrac{1}{2}  \\ \hline
\textcolor{red}{\psi^2}  & -\tfrac{1}{2} &0 & 0&0 & -1& 1&0 &0& \tfrac{1}{2}  \\ \hline
\boldsymbol{\textcolor{red}{\psi^3}} &-\tfrac{1}{2}  & 0& 0& 0& 0&-1 & 1&0& \tfrac{1}{2}  \\ \hline
\boldsymbol{\psi^4} &-\tfrac{1}{2}  &0 &0 & 0& 0&0 &-1 &-\tfrac{1}{2}& -\tfrac{1}{2}  \\ \hline
\end{array}
\end{equation}
where the \textbf{bold} letters denote those fields selected by the physical state condition \eqref{eq:WedgeSubset} and the meaning of the \textcolor{red}{red} coloring will be apparent momentarily.

\medskip

Now we would like to restrict to a subset of all the free fields in \eqref{eq:YZlist} such that their $\mathcal{C}$-charge neutral bilinears satisfy the Schur condition \eqref{eq:SchurCondition}.
(We will impose further physical state conditions afterwards.)
This condition selects a subset of the two lists in \eqref{eq:YZlist}:
\begin{equation}\label{eq:SchurSubset}
\begin{aligned}
Y_I \,\, \supset \,\,
{}^{\textrm{S}}Y_I
&=(\mu^{\dagger}_2
\,,\,
\lambda^{\dagger}_1
\,,\, 
\psi^{\dagger}_2 
\,,\,
 \psi^{\dagger}_3) 
\,,  \\
Z^I \,\, \supset \,\, 
{}^{\textrm{S}}Z^I
&=(\lambda^2
\,,\,
\mu^1
\,,\, 
\psi^2 
\,,\, 
\psi^3) \,,
\end{aligned}
\end{equation}
which we will call the ``worldsheet Schur subsector" and which is  colored \textcolor{red}{red} in the table \eqref{tabel:FreeFieldCharge}.
We emphasize that the 4D spectrum that corresponds to this worldsheet Schur subsector is much larger than the Schur subsector of the 4D theory: only the short multiplets (see below) in this  worldsheet Schur subsector reproduces the Schur subsector of the 4D theory.

\medskip

One can then check directly that the collection of free fields in \eqref{eq:SchurSubset} generate the current algebra $\mathfrak{u}(1,1|2)_1$.
And then imposing the $\mathcal{C}=0$ constraint\footnote{Or more precisely, we impose $\mathcal{C}_n=0$ for $n\geq0$; and the $\mathcal{C}_{-n}$ descendants are then null and are naturally quotiented out.} takes us from $\mathfrak{u}(1,1|2)_1$ to $\mathfrak{psu}(1,1|2)_1$.
Indeed, one can make the following identification of the free fields in \eqref{eq:SchurSubset} with the free fields in the worldsheet theory of \ASTT, listed in \eqref{eq:FreeFieldAdS3}:
\begin{equation}\label{eq:matchAdSfields}
\begin{aligned}
&(\mu^{\dagger}_2
\,,\, 
\lambda^{\dagger}_1)
=(\xi^{+}
\,,\, 
\xi^{-})
\,, \qquad \quad
(\mu^{1}
\,,\, 
-\lambda^2)
=(\eta^{+}
\,,\, 
\eta^{-})\,,\\
&(\psi^{\dagger}_3
\,,\, 
\psi^{\dagger}_2)
=(\phi^{+}
\,,\, 
\phi^{-})
\,,\qquad  \quad
(-\psi^{2}
\,,\, 
\psi^{3})=(\chi^{+}
\,,\, 
\chi^{-})\,.
\end{aligned}
\end{equation}
Then since the fields in \eqref{eq:FreeFieldAdS3} generate the current algebra $\mathfrak{psu}(1,1|2)_1$ under the $\mathcal{C}=0$ constraint, so do the fields in the subset \eqref{eq:SchurSubset}.

\subsubsection{Imposing physical state condition}

Now that we have the fields \eqref{eq:SchurSubset} that are selected out by the Schur condition \eqref{eq:SchurCondition}, we impose the physical state conditions \eqref{eq:WedgeCondition}, \eqref{eq:L0constraint}, and \eqref{eq:Cconstraint}. 
In particular, we need to check that after imposing the physical state conditions, we are not left with an empty set. 

\medskip

First, we impose the wedge condition \eqref{eq:WedgeSubset} and \eqref{eq:WedgeCondition} on the worldsheet Schur subset \eqref{eq:SchurSubset}.
This first selects a subset of \eqref{eq:SchurSubset}:
\begin{equation}\label{eq:SchurWedge1}
\begin{aligned}
Y_I 
\,\,\supset \,\,
{}^{\textrm{S}}Y_I
&
\,\,\supset\,\, 
{}^{\textrm{S},\vee}Y_I =
(\mu^{\dagger}_2
\,,  \,
\psi^{\dagger}_2 )
\,, \qquad \\
Z^I 
\,\,\supset \,\,
{}^{\textrm{S}}Z^I
&
\,\,\supset\,\,
{}^{\textrm{S},\vee}Z^I=
(\mu^1
\,, \,
\psi^3)\,,
\end{aligned}
\end{equation}
which can be loosely called the ``physical subset of worldsheet Schur fields" and   are denoted by \textcolor{red}{\textbf{bold red}} letters in \eqref{tabel:FreeFieldCharge}, and then further restrict to the ``worldsheet Schur wedge-modes", defined as
\begin{equation}\label{eq:SchurWedge2}
(\mu^\dagger_2)_r 
\ , \quad 
(\mu^{1})_r 
\ , \quad   
(\psi^\dagger_{2})_r  
\ ,  
\quad (\psi^{3})_r 
\qquad \textrm{with} \quad 
-\tfrac{w-1}{2} \leq r \leq \tfrac{w-1}{2}\,.
\end{equation}

Next, recall that for the full theory, in the $w$-spectrally-flowed sector, imposing the Virasoro constraint \eqref{eq:L0constraint} and the Central constraint \eqref{eq:Cconstraint} together with the wedge mode constraint \eqref{eq:WedgeSubset} and \eqref{eq:WedgeCondition} amounts to considering only the $\mathcal{C}$-charge neutral DDF-like operators \eqref{eq:DDF} that are bilinears of the form $({}^{\vee}Y_I{}^{\vee}Z^J)$, where the modes of both ${}^{\vee}Y_I$ and ${}^{\vee}Z^J$ are restricted to their wedge modes. 
Furthermore, we require that the total mode number is zero mod $w$.

Therefore, within the Schur subsector \eqref{eq:SchurSubset}, after imposing the physical state condition \eqref{eq:WedgeSubset}, \eqref{eq:WedgeCondition}, \eqref{eq:L0constraint}, and \eqref{eq:Cconstraint}, we are left with $\mathcal{C}$-charge neutral  DDF-like operators \begin{equation}\label{eq:DDFSchur}
{}^{\textrm{S}}(S_I^{\,\,J})_m=\sum^{\frac{w-1}{2}}_{r=m-\frac{w-1}{2}}({}^{\textrm{S},\vee}Y_I)_r({}^{\textrm{S},\vee}Z^J)_{m-r}
\quad \textrm{with}\quad
m=0,1,\dots, w-1\,,
\end{equation}
for each $w$-spectrally-flowed sector, where ${}^{\textrm{S},\vee}Y_I$ and ${}^{\textrm{S},\vee}Z^J$ are from the list \eqref{eq:SchurWedge1} and their modes numbers are inside the wedge \eqref{eq:SchurWedge2}.
There are only four such $\mathcal{C}$-neutral bilinears that one can build from the physical subset of worldsheet Schur fields in \eqref{eq:SchurWedge1}. 
We list their charges below
\begin{equation}\label{tabel:C0bilinears}
\begin{array}{|c|c|c|c|c|c|c|c|c|c|}
\hline
&\,\, \,\,
\mathcal{C}
\,\,\,\,&\,\,\,\,
E 
\,\,\,\,&\,\,\,\,
j_1
\,\,\,\,&\,\,\,\,
j_2 
\,\,\,\,&\,\,\,\,
R 
\,\,\,\,&\,\,\,\,
r\,\,\,\,
\\ \hline
\boldsymbol{\textcolor{red}{\mu^{\dagger}_{2} \mu^{1}}} & 0& 1& \tfrac{1}{2} & \tfrac{1}{2} &0& 0\\ \hline 
\boldsymbol{\textcolor{red}{\mu^{\dagger}_{2} \psi^{3}}} & 0&  \tfrac{1}{2}& \tfrac{1}{2} & 0&0&  \tfrac{1}{2}\\ \hline 
\boldsymbol{\textcolor{red}{\psi^{\dagger}_{2} \mu^{1}}} & 0&  \tfrac{1}{2} & 0 & \tfrac{1}{2}&0& - \tfrac{1}{2}\\ \hline 
\boldsymbol{\textcolor{red}{\psi^\dagger_2 \psi^{3}}}&0 &0 & 0& 0&0&0 \\ \hline
\end{array}
\end{equation}
which obey the Schur conditions \eqref{eq:SchurCondition} by design.

Finally, applying the products:
\begin{equation}\label{eq:L0constraintSchur}
\prod_i{}^{\textrm{S}}(S_I^{\,\,J})_{m_i}
\quad \textrm{with}\quad  
\sum_i m_i=0 \quad \textrm{mod} \,\,w\,,
\end{equation} 
on the ground state $|0\rangle_w$, we obtain all the physical states in the $w$-spectrally-flowed sector of the Schur subsector. 

\subsection{Reproducing Schur index}

Now we show that the worldsheet Schur wedge modes in \eqref{eq:SchurWedge2} reproduce the Schur index of 4D $\mathcal{N}=4$ SYM.

\subsubsection{Reproducing single-letter index from \texorpdfstring{$w=1$}{w=1} sector}

Let's first consider the $w=1$ spectrally-flowed sector, which should reproduce the single-letter Schur index of 4D $\mathcal{N}=4$ SYM.
The vacuum is the RR-vacuum, with conformal dimension 
\begin{equation}\label{eq:RRvach}
|0\rangle_{w=1}: \qquad h=\tfrac{1}{2} \,.
\end{equation}
Now we compute the characters of (the physical spectrum of) the $w=1$ Schur subsector of the worldsheet theory  by applying the $\mathcal{C}$-neutral bilinears composed of the Schur wedge-modes in  \eqref{eq:SchurWedge2}, for $w=1$, repeatedly on the vacuum.
From \eqref{eq:SchurWedge2}, we can see that for $w=1$, the wedge modes are simply the zero modes.
Then due to the commutation relations of the field in \eqref{eq:SchurWedge1}, we can apply the bilinear $\mu^{\dagger}_{2,0}\mu^1_0$ infinitely many times, but the other three bilinears, namely
\begin{equation}
\mu^{\dagger}_{2,0}\psi^{3}_0
\,, \qquad 
\psi^{\dagger}_{2,0}\mu^{1}_0
\,,\qquad 
\psi^{\dagger}_{2,0}\psi^{3}_{0} 
\end{equation}
at most once.

Therefore, there are four types of  contributions to the character:
\begin{equation}\label{eq:4contributionw=1}
\begin{aligned}
&(\mu^{\dagger}_{2,0}\mu^1_0)^{n} |0\rangle_1
\,, \qquad 
(\mu^{\dagger}_{2,0}\mu^1_0)^{n} (\psi^{\dagger}_{2,0}\psi^{3}_{0})|0\rangle_1
: \qquad 
\frac{q^{\tfrac{1}{2}}}{1-q} \chi_{\tfrac{1}{2}}(a) \\
&(\mu^{\dagger}_{2,0}\mu^1_0)^{n} (\mu^{\dagger}_{2,0}\psi^{3}_0) |0\rangle_1
\,, \qquad 
(\mu^{\dagger}_{2,0}\mu^1_0)^{n} (\psi^{\dagger}_{2,0}\mu^{1}_0)|0\rangle_1
: \qquad 
\frac{2q}{1-q}  \,,
\end{aligned}
\end{equation}
where $n=0,1,\dots,\infty$, and $a$ is the chemical potential corresponding to $\frac{R_2}{2}$, which is the Cartan of $\mathfrak{su}(2)_F$  (see \eqref{eq:indexN4vec}), and the charges of the fermions can be read off from Table \eqref{tabel:FreeFieldCharge}.
In total, the  character of the Schur subsector of the RR vacuum is therefore\footnote{It is denoted as $Z_{\myng(1)}(a,q)$ in \cite{Gaberdiel:2022iot}.}
\begin{equation}\label{eq:SchurRRcharacter}
\begin{aligned}
\pfWSw^{(1)}(a,q)&=\frac{q^{\frac{1}{2}}}{1-q}\left(\chi_{\frac{1}{2}}(a)+2 \, q^{\frac{1}{2}} \right):=\pfWSw(a,q)\,.
\end{aligned}
\end{equation}

\medskip

To compute the corresponding  index, note that the fermionic states (in the second line of \eqref{eq:4contributionw=1}) acquire an additional minus sign; therefore the index is
\begin{equation}\label{eq:SchurRRIndex}
\begin{aligned}
\IWSw^{(1)}(a,q)&=\frac{q^{\frac{1}{2}}}{1-q}\left( \chi_{\frac{1}{2}}(a)-2 \, q^{\frac{1}{2}} \right)=:\IWSw(a,q)\,.
\end{aligned}
\end{equation}
This then reproduces the single-letter Schur index of the 4D $\mathcal{N}=4$ vector-multiplet \eqref{eq:indexVecN4Schur}:
\begin{equation}\label{eq:MatchSchurw=1}
\IWSw(a,q) = \IFsl^{\textrm{Schur}}_{\textrm{vec}}(a,q)\,,
\end{equation}
where the l.h.s.\ and r.h.s\ are from the worldsheet and spacetime computations, respectively. 

\medskip

Note that in both the character \eqref{eq:SchurRRcharacter} and the index \eqref{eq:SchurRRIndex}, the coefficients for the leading terms $q^{\frac{1}{2}}$  are positive, even though it is a half-integer mode; this is due to the fact that the RR vacuum $|0\rangle_1$ has conformal dimension $h=\frac{1}{2}$ (see \eqref{eq:RRvach}) but is nevertheless ``bosonic". 
Namely, to go from the character to the index, instead of flipping the signs of all terms $\sim \mathcal{O}(q^{n+\frac{1}{2}})$, one should apply the following operation
\begin{equation}\label{eq:ChaToInd}
\pfWSw(a,q)
\quad \longrightarrow\quad \frac{q^{\frac{1}{2}}}{1-q}\cdot \left(\left(\frac{1-q}{q^{\frac{1}{2}}}\cdot\pfWSw(a,q)\right)\big|_{q^{\frac{1}{2}}\rightarrow(-1)q^{\frac{1}{2}}}\right)  
= \,\,
\IWSw(a,q)\,.
\end{equation}

\subsubsection{Including all spectrally-flowed sectors}

Once we have the spectrum of the $w=1$ sector, those of the general $w$ spectrally-flowed sectors can be derived using the argument reviewed in Section.\ \ref{ssec:SpectralFlowAdS5}. 
First, we find that the character in the $w$-spectrally-flowed sector \eqref{eq:ZwCyc} can be rewritten into a more convenient form
\begin{equation}
 \pfWS^{(w)}_{\textrm{\ASF}}(\mathfrak{q}) =
\frac{1}{w} \left( \sum_{d|w, d \textrm{ odd}}\, \phi(d)\, \pfWS(\mathfrak{q}^d)^{\frac{w}{d}} + \sum_{d|w, d \textrm{ even}}\, \phi(d)\, \widetilde{\pfWS}(\mathfrak{q}^d)^{\frac{w}{d}}\right)\,.
\end{equation}
Specializing to the worldsheet Schur subsector, the character of the worldsheet Schur wedge-modes \eqref{eq:SchurWedge2} in the $w$-spectrally-flowed sector is then \begin{equation}\label{eq:WSzw}
\pfWSsl^{(w)}(a,q)\equiv 
\frac{1}{w} \left( \sum_{d|w, d \textrm{ odd}}\, \phi(d)\, \pfWSsl(a^d,q^d)^{\frac{w}{d}} + \sum_{d|w, d \textrm{ even}}\, \phi(d)\, \IWSsl(a^d,q^d)^{\frac{w}{d}}\right)\,,
\end{equation}
and the full character from the worldsheet Schur wedge-modes \eqref{eq:SchurWedge2} is
\begin{equation}\label{eq:wsSchurCharacter}
\pfWS^{\textrm{ws, Schur}}_{U(N)}(a,q)
=\sum^{\infty}_{w=1}\pfWSsl^{(w)}(a,q)\,.
\end{equation}

\medskip

Let us now compute the corresponding index. 
For each $w$-spectrally-flowed sector, similar to the $w=1$ case, we cannot just flip the signs for each half-integer term in the character, since the ground state $|0\rangle_w$ of the $w$-spectrally-flowed sector has conformal dimension $h=\frac{w}{2}$ but is always ``bosonic".
To obtain the index $\IWSsl^{(w)}(a,q)$ from the character $\pfWSsl^{(w)}(a,q)$, we should apply the operation similar to \eqref{eq:ChaToInd} on each term in $\pfWSsl^{(w)}(a,q)$:
\begin{equation}\label{eq:ChaToIndw}
\begin{aligned}
d \textrm{ odd}: \quad 
&\pfWSw(a^d,q^d)
\,\, \longrightarrow\,\, 
\frac{q^{\frac{d}{2}}}{1-q^d}\cdot \left(\left(\frac{1-q^d}{q^{\frac{d}{2}}}\cdot\pfWSw(a^d,q^d)\right)\big|_{q^{\frac{1}{2}}\rightarrow(-1)q^{\frac{1}{2}}}\right)  
= \,\,
\IWSw(a^d,q^d)\,; \\
d \textrm{ even}: \quad 
&\IWSw(a^d,q^d)
\,\, \longrightarrow\,\, 
\frac{q^{\frac{d}{2}}}{1-q^d}\cdot \left(\left(\frac{1-q^d}{q^{\frac{d}{2}}}\cdot\IWSw(a^d,q^d)\right)\big|_{q^{\frac{1}{2}}\rightarrow(-1)q^{\frac{1}{2}}}\right)  
= \,\,
\IWSw(a^d,q^d)\,. \\
\end{aligned}
\end{equation}
Therefore, the  index $\IWSsl^{(w)}(a,q)$ from the $w$-spectrally-flowed sector is 
\begin{equation}\label{eq:WSiw}
\IWSsl^{(w)}(a,q)\equiv \frac{1}{w} \sum_{d|w}\, \phi(d)\, \IWSsl(a^d,q^d)^{\frac{w}{d}}\,.
\end{equation}
And the total index from the worldsheet Schur wedge-modes \eqref{eq:SchurWedge2} is
\begin{equation}\label{eq:wsSchurIndex}
\IWS^{\textrm{ws, Schur}}_{U(N)}(a,q)
=\sum^{\infty}_{w=1}\IWSsl^{(w)}(a,q)\,,
\end{equation}
and for the $SU(N)$ theory the summation starts from $w=2$.
With the identification \eqref{eq:MatchSchurw=1}, this then reproduces the single-particle Schur index of $\mathcal{N}=4$ SYM \eqref{eq:SPIndex4Dsec3}.

\section{Chiral algebra from worldsheet}

In the previous section, we showed that the physical states in the Schur subsector of the worldsheet theory are generated by the $\mathcal{C}$-neutral bilinears of the wedge modes of the free fields in the Schur subsector. 
In this section, we study the algebra generated by these physical states and then show that it contains the chiral algebra as its subalgebra that is generated by the short multiplets of the $\mathcal{N}=4$ superconformal algebra.

\subsection{Worldsheet Schur algebra and symmetry enhancement at free point}

The physical Schur states we identified in the previous section generate a large symmetry algebra, which we will call the ``worldsheet Schur algebra":
\begin{equation}\label{eq:worldsheetSchurAlgebra}
\begin{aligned}
&\textrm{worldsheet Schur algebra}:  \textrm{the $\mathcal{N}=4$ $\mathcal{W}_{\infty}$ algebra  generated by}\\
&\qquad\qquad\,\,\quad  \qquad \qquad \qquad \textrm{ the physical states as counted by \eqref{eq:wsSchurCharacter}}\,.
\end{aligned}
\end{equation}
We will give the full (multi-particle) vacuum character later, which comes from all the products of the single-particle physical states and is not directly part of the worldsheet theory.
We emphasize that this worldsheet Schur algebra is not to be confused with the chiral algebra:  the latter is a small subalgebra of the former that is only generated by short multiplets (see below).

It is a very large algebra: in particular, it has an  $\mathcal{N}=4$ even spin $\mathcal{W}_{\infty}$ subalgebra, which is generated by the Schur physical states from the $w=2$ spectrally-flowed sector, whose character has the expansion:\footnote{
Note that this is similar to the spectrum of the $\mathcal{N}=4$ $\mathcal{W}_{\infty}$ algebra that arises as the boundary symmetry of the $\mathcal{N}=4$ Vasiliev higher-spin gravity in AdS$_3$, except that the latter has one long multiplet for each positive integer spin, not just the even ones, see \cite{Gaberdiel:2013vva}.}
\begin{equation}
\pfWSw^{(2)}(a,q) = \ChNf_{{\rm{vac}}} (a,q)  + \sum_{n=1}^{\infty} \ChLoNf_{h=2n,0} (a,q)  \ ,
\end{equation}
where $\ChNf_{{\rm{vac}}} (a,q)$ is the vacuum character of the $\mathcal{N}=4$ superconformal algebra
\begin{equation}\label{eq:N4chaVac}
\ChNf_{{\rm{vac}}} (a,q)  =  \frac{q}{(1-q)} \left(
 \chi_1(a) 
+ 2 \, q^{1/2} \, \chi_{\frac{1}{2}}(a)
 + q\, \chi_0(a) \right)  \ , 
\end{equation}
where the lowest component is the adjoint of the $\mathfrak{su}(2)_R$ (the R-symmetry of the 2D $\mathcal{N}=4$ superconformal algebra);\footnote{Note that the $\mathfrak{su}(2)_R$ in this section refers to the R-symmetry of the 2D $\mathcal{N}=4$ superconformal symmetry, which corresponds to the $\mathfrak{su}(2)_F$ flavor symmetry of 4D $\mathcal{N}=4$ SYM, by the argument of \cite{Beem:2013sza} --- it is not to be confused with the $\mathfrak{su}(2)_R$ R-symmetry from the $\mathcal{N}=2$ subalgebra of the 4D theory.}
and $\ChLoNf_{h,0} (a, q) $ is the character of the long multiplet of the $\mathcal{N}=4$ superconformal algebra whose lowest component is an $\mathfrak{su}(2)_R$ singlet:
\begin{equation}\label{eq:N4chaLong}
\ChLoNf_{h,0}  (a, q) = \frac{q^h}{(1-q)} \left( 
1  +  2 \, q^{1/2}\, \chi_{1/2}(a) + q \,(\chi_1(a) + 3) +
2 \, q^{3/2}\,  \chi_{1/2}(a) + q^2 \right) \ .
\end{equation}
The fact that the decomposition is in terms of representations of the $\mathcal{N}=4$ superconformal algebra, which is  the symmetry algebra of the boundary dual CFT of the \ASTM, will be explained later in Sec.\ \ref{ssec:AST}.

\medskip

Similarly, the index of the algebra \eqref{eq:worldsheetSchurAlgebra} is \eqref{eq:wsSchurIndex}, which by construction reproduces the single-particle Schur index of the 4D theory:
\begin{equation}\label{eq:MatchSTWSSchurInd}
\mathcal{I}^{\textrm{st, sp, Schur}}_{U(N)}(a,q)
=\IWS^{\textrm{ws, Schur}}_{U(N)}(a,q)
=\sum^{\infty}_{w=1}\IWSsl^{(w)}(a,q)\,,
\end{equation}
and again for the $SU(N)$ theory the summation starts from $w=2$.
However, the chiral algebra that we want to reproduce from the worldsheet should only be a (small) subalgebra of the worldsheet Schur algebra \eqref{eq:worldsheetSchurAlgebra}, since the latter only exists at the free point of the 4D theory, whereas the chiral algebra is independent of the coupling. 
Namely, the character of \eqref{eq:worldsheetSchurAlgebra} is much larger than that of the chiral algebra.
 
\subsection{Chiral algebra from BPS subsector of Schur subsector}

In order to extract the chiral algebra from the worldsheet, which only describes the free point of the 4D theory, we need to focus on the part of the spectrum that does not get lifted once we turn on the coupling.
This suggests us to focus on the ``BPS sector" of worldsheet Schur subsector, given by \eqref{eq:SchurWedge2}. 
Namely, the character of the chiral algebra should allow a decomposition in terms of characters of only short multiplets of the $\mathcal{N}=4$ superconformal algebra. 

\medskip

Let us consider a short multiple of the $\mathcal{N}=4$ superconformal  algebra whose lowest component is the spin-$s$ representation of $\mathfrak{su}(2)_R$, with 
\begin{equation}
\textrm{conformal dimension }h = \textrm{spin }s\,.
\end{equation}
The character of such a short multiplet is
\begin{equation}\label{eq:BPSCharacterh}
\begin{aligned}
&\ChShNf_{h\geq 1}(a,q)\equiv \frac{q^h}{1-q}\left(\chi_{h}(a)
+2\, q^{\frac{1}{2}}\, \chi_{h-\frac{1}{2}}(a)
+q\, \chi_{h-1}(a)\right)\,,\\
&\ChShNf_{\frac{1}{2}}(a,q)\equiv \frac{q^{\frac{1}{2}}}{1-q}\left(\chi_{\frac{1}{2}}(a)
+2\, q^{\frac{1}{2}}\right) \,,\\
%&\ChShNf_{0}(a,q)\equiv 1\,,\\
\end{aligned}
\end{equation}
where $\chi_s(a)$ is the character of the spin-$s$ representation of  $\mathfrak{su}(2)_{R}$; $h$ can take all positive integer and half-integer values.
The vacuum representation \eqref{eq:N4chaVac} is such a short multiplet with $h=s=1$, generated by the $\mathcal{N}=4$ superconformal generators
\begin{equation}
J^{a}_{-1} 
\,, \quad 
G^{\pm}_{-\frac{3}{2}}
\,, \quad
\bar{G}^{\pm}_{-\frac{3}{2}}
\,, \quad 
L_{-2}\,,
\end{equation}
acting on the vacuum state $|0\rangle$. 
The correspondingly index is
\begin{equation}\label{eq:BPSindexh}
\begin{aligned}
&\IShNf_{h\geq \frac{1}{2}}(a,q)\equiv \frac{q^h}{1-q}\left(\chi_{h}(a)
-2\, q^{\frac{1}{2}}\, \chi_{h-\frac{1}{2}}(a)
+q\, \chi_{h-1}(a)\right)\,,\\
&\IShNf_{\frac{1}{2}}(a,q)\equiv \frac{q^{\frac{1}{2}}}{1-q}\left(\chi_{\frac{1}{2}}(a)
-2\, q^{\frac{1}{2}}\right)\,. \\
%&\IShNf_{0}(a,q)\equiv 1\,.
\end{aligned}
\end{equation}
Note that the bottom component always appears with a ``+" sign, even for half-integer $h$. 

\medskip

In comparison, the character of a long multiplet of the $\mathcal{N}=4$ superconformal  algebra whose lowest component is the spin-$s$ representation of $\mathfrak{su}(2)_R$ is 
\begin{equation}\label{eq:LongCharacterhs}
\begin{aligned}
&\ChLoNf_{h,s\geq 1}(a,q)\equiv \frac{q^h}{1-q}\big(
\chi_{s}(a)
+2\, q^{\frac{1}{2}} \, (\chi_{s-\frac{1}{2}}(a)+\chi_{s+\frac{1}{2}}(a))\\
&\qquad\qquad+q\, (\chi_{s-1}(a)+4\chi_{s}(a)+\chi_{s+1}(a)) 
+2\,q^{\frac{3}{2}}\, (\chi_{s-\frac{1}{2}}(a)+\chi_{s+\frac{1}{2}}(a))
+q^2\, \chi_{s}(a) 
\big)\,, \\
&\ChLoNf_{h,\frac{1}{2}}(a,q)\equiv \frac{q^h}{1-q}\big(
\chi_{\frac{1}{2}}(a)
+2\, q^{\frac{1}{2}} \, (1+\chi_{1}(a))\\
&\qquad \qquad +q\, (4\chi_{\frac{1}{2}}(a)+\chi_{\frac{3}{2}}(a)) 
+2\,q^{\frac{3}{2}}\, (1+\chi_{1}(a))
+q^2\, \chi_{\frac{1}{2}}(a) 
\big)\,, \\
&\ChLoNf_{h,0}(a,q)\equiv  \frac{q^h}{(1-q)} \left( 
1  +  2 \, q^{1/2}\, \chi_{1/2}(a) + q \,(\chi_1(a) + 3) +
2 \, q^{3/2}\,  \chi_{1/2}(a) + q^2 \right)  \,,
\end{aligned}
\end{equation}
where the last one $\ChLoNf_{h,0}(a,q)$ was already given earlier in \eqref{eq:N4chaLong}, and we list it here to show the shortening condition of $\ChLoNf_{h,s}(a,q)$ for small $s$'s.
Then flipping the signs of the $q^{\frac{1}{2}}$ and $q^{\frac{3}{2}}$ terms inside the brackets gives their corresponding indices $\ILoNf_{h,s}(a,q)$, which will not be important in this paper.

\medskip

The main observation is that  the index \eqref{eq:wsSchurIndex} of the algebra \eqref{eq:worldsheetSchurAlgebra}, which reproduces the single-particle Schur index of the 4D theory, has the following decomposition in terms of the indices \eqref{eq:BPSindexh}:\footnote{
Recall that the single-particle Schur index for the $SU(N)$ theory is given by $\IFsp^{\textrm{Schur}, SU(N)}(a,q)=\IFsp^{\textrm{Schur}, U(N)}(a,q)-\IFsl^{\textrm{Schur}}_{\textrm{vec}}(a,q)$, where $\IFsl^{\textrm{Schur}}_{\textrm{vec}}(a,q)$ is the single-letter Schur index.} 
\begin{equation}\label{eq:ZUNsingleBPSExp}
\boxed{
\mathcal{I}^{\textrm{st, sp, Schur}}_{U(N)}(a,q)
=\IWS^{\textrm{ws, Schur}}_{U(N)}(a,q) 
=\sum^{\infty}_{w=1} \IShNf_{\frac{w}{2}}(a,q)\,.
}
\end{equation}
The fact that only the indices of the short multiplets appear is reassuring since we are considering the spacetime Schur index, which counts the BPS spectrum.

\smallskip

Now let us compare the decompositions  \eqref{eq:MatchSTWSSchurInd} and \eqref{eq:ZUNsingleBPSExp}:
\begin{equation}\label{eq:twodecompositions}
\mathcal{I}^{\textrm{st, sp, Schur}}_{U(N)}(a,q)
=\IWS^{\textrm{ws, Schur}}_{U(N)}(a,q) 
=\sum^{\infty}_{w=1}\IWSsl^{(w)}(a,q)
=\sum^{\infty}_{w=1} \IShNf_{\frac{w}{2}}(a,q)\,.
\end{equation}
For this identity on indices to exist, there are lots of cancellations involved.  
First, we emphasize that the cancellation happens across different $w$; namely 
\begin{equation}
%	\sum_{\{h,s\}} n^{(w)}_{h,s} \ILoNf_{h,s}(a,q)
%	\neq 0
%	\qquad \longrightarrow \qquad 
	\IWSsl^{(w)}(a,q)\neq  \IShNf_{\frac{w}{2}}(a,q)\,,
\end{equation} 
for $w\geq 2$, 
which makes the identity between the two decompositions in \eqref{eq:twodecompositions} rather non-trivial.
Furthermore, we note that the spectrum underlying $\IWSsl^{(w)}(a,q)$ is much larger than the spectrum underlying $ \IShNf_{\frac{w}{2}}(a,q)$.
To see this, one can compare the characters\footnote{
It is easier to do this comparison after multiplying both $\pfWSw^{(w)}(a,q)$ and $\ChShNf_{h=\frac{w}{2}}(a,q)$ by the factor $(1-q)$, since the factor $(1-q)$ only signifies that we are including all the modes that are from derivatives.} $\pfWSsl^{(w)}(a,q)$ and $\ChShNf_{\frac{w}{2}}(a,q)$.
It is easy to check that the $\pfWSw^{(w)}(a,q)$ ``contains" $\ChShNf_{h=\frac{w}{2}}$ as its leading terms, 
namely, the difference between the two characters 
\begin{equation}
\pfWSw^{(w)}(a,q) -\ChShNf_{h=\frac{w}{2}}(a,q)
\end{equation}
contains \textit{only positive terms}.
Moreover, we find that 
\begin{equation}\label{eq:difference}
\pfWSw^{(w)}(a,q) -\ChShNf_{h=\frac{w}{2}}(a,q) 
= \sum_{\{h,s\}} n^{(w)}_{h,s}\,
\ChLoNf_{h,s}(a,q)\,,
\end{equation}
with \textit{non-negative} integer coefficients $n^{(w)}_{h,s} \in \mathbb{N}_0$. 
For example, for the first few $w$, the differences in \eqref{eq:difference} are
\begin{equation}
\begin{aligned}
w=1: \qquad& 0\\
w=2: \qquad& \sum^{\infty}_{n=1}\ChLoNf_{2n,0}(a,q)\\
w=3:\qquad &
2\ChLoNf_{2,0}(a,q)
+\ChLoNf_{\frac{5}{2},\frac{1}{2}}(a,q)
+2\ChLoNf_{\frac{7}{2},\frac{1}{2}}(a,q)
+2\ChLoNf_{4,0}(a,q)\\
&+\ChLoNf_{\frac{9}{2},\frac{1}{2}}(a,q)+4\ChLoNf_{5,0}(a,q)+\dots \\
w=4:\qquad &
\ChLoNf_{2,0}(a,q)
+2\ChLoNf_{\frac{5}{2},\frac{1}{2}}(a,q)
+2\ChLoNf_{3,1}(a,q)
+\ChLoNf_{3,0}(a,q)\\
&+4\ChLoNf_{\frac{7}{2},\frac{1}{2}}(a,q)
+2\ChLoNf_{4,1}(a,q)
+7\ChLoNf_{4,0}(a,q)+\dots \\
\end{aligned}
\end{equation}

\medskip

Since the spectrum on the r.h.s.\ are all from long multiplets, as we move away from the free point, they are expected to be lifted.
Therefore the character for the worldsheet spectrum that generates the chiral algebra, which is independent of the coupling, should be
\begin{equation}\label{eq:ZUNsingleBPSExpCh}
\pfWS^{\textrm{ws, Schur}}_{U(N)}(a,q) 
=\sum^{\infty}_{w=1} \ChShNf_{\frac{w}{2}}(a,q)\,.
\end{equation}
Namely, the chiral algebra is an $\mathcal{N}=4$ $\mathcal{W}_{\infty}$ algebra generated by the modes captured by the short $\mathcal{N}=4$ characters $\ChShNf_{h}(a,q)$, defined  in \eqref{eq:BPSCharacterh}, with $h\in \frac{1}{2} \mathbb{N}$.
Since $\ChShNf_{h}(a,q)$ is a character, with only positive coefficients, it uniquely determines the worldsheet content that generates the chiral algebra, and hence the spin-content of the chiral algebra,  at general coupling.

This is consistent with the picture that at the free point, for which the worldsheet theory is valid, the symmetry is vastly enhanced, captured by the bigger characters $\pfWSw^{(w)}(a,q)$.
Away from the free points, most of the spectrum gets lifted and the remaining generators are described by $\ChShNf_{h=\frac{w}{2}}(a,q)$.

Finally, the decomposition \eqref{eq:ZUNsingleBPSExpCh} matches nicely with Conjecture 3 of \cite{Beem:2013sza} for $G=U(N)$ and $N\rightarrow\infty$. And each $\ChShNf_{h\geq 1}(a,q)$ in \eqref{eq:ZUNsingleBPSExpCh} (defined in \eqref{eq:BPSCharacterh}) contains 4 terms and they have conformal dimension and spin
\begin{equation}
(h,s)\,,
\quad (h+\frac{1}{2},s-\frac{1}{2})	\,,
\quad (h+\frac{1}{2},s-\frac{1}{2})	\,,
\quad (h+1,s-1)	\,,
\end{equation}
respectively, corresponding to the 4 families of operators in \cite{Beem:2013sza,Costello:2018zrm}.

\subsection{Full character and index of vacuum representation of chiral algebra}

Finally, we consider the multi-particling of the  worldsheet character  \eqref{eq:ZUNsingleBPSExpCh} to derive the full character and the index of the vacuum representation of the chiral algebra.
As we will see, there are a lot of cancellations in the index, which reproduces the full large-$N$ Schur index \eqref{eq:IndexLargeNSchurUN} from the spacetime.

\medskip

First of all, each factor in \eqref{eq:ZUNsingleBPSExpCh} contributes
\begin{equation}
q^h \chi_{j}(a) 
\qquad \longrightarrow\qquad
\begin{cases}
\begin{aligned}
&\prod^{\infty}_{n=h}\prod^{j}_{m=-j} 
 \frac{1}{(1-a^m q^n)}\qquad \textrm{bosonic}
\\
&\prod^{\infty}_{n=h}\prod^{j}_{m=-j} 
 (1\pm a^m q^n)\qquad \,\,\textrm{fermionic} \,,
\end{aligned}
\end{cases}
\end{equation}
where for the fermionic case in the second line, ``$+$" is for the character whereas ``$-$" is for the index.

\medskip

Therefore, to the full character  of the vacuum representation, each  $\ChShNf_{\frac{w}{2}}(a,q)$ (defined in \eqref{eq:BPSCharacterh}) in the sum \eqref{eq:ZUNsingleBPSExpCh} contributes
\begin{equation}
\begin{aligned}
\ChShNf_{\frac{w}{2}}(a,q)
\quad&\longrightarrow\quad \mathfrak{Z}^{\mathcal{N}=4,\textrm{short}}_{\frac{w}{2}}(a,q)=
\prod^{\infty}_{n=\frac{w}{2}}\prod^{\frac{w}{2}}_{m=-\frac{w}{2}} 
 \frac{1}{(1-a^m q^n)}\\
&
\cdot \prod^{\infty}_{n=\frac{w+1}{2}}\prod^{\frac{w-1}{2}}_{m=-\frac{w-1}{2}} 
 (1+ a^m q^n)^2
\cdot \prod^{\infty}_{n=\frac{w}{2}+1}\prod^{\frac{w}{2}-1}_{m=-\frac{w}{2}+1} 
 \frac{1}{(1-a^m q^n)} \,.
\end{aligned}
\end{equation}
Taking all the spectrally-flowed sectors into account, the full character of the vacuum representation of the chiral algebra is then  
\begin{equation}\label{}
\begin{aligned}
\mathfrak{Z}^{\textrm{ws, Schur}}_{U(N)}(a,q)
= \prod^{\infty}_{w=1} \mathfrak{Z}^{\mathcal{N}=4,\textrm{short}}_{\frac{w}{2}}(a,q)\,.
\end{aligned}
\end{equation}
There is no cancellation between numerators and denominators due to the sign difference. 

\medskip

On the other hand, to the full index  of the vacuum representation, each  \\
$\IShNf_{\frac{w}{2}}(a,q)$ (defined in \eqref{eq:BPSindexh}) in the sum \eqref{eq:ZUNsingleBPSExp} contributes
\begin{equation}\label{eq:FullIndexWSw}
\begin{aligned}
\IShNf_{\frac{w}{2}}(a,q)
\quad&\longrightarrow\quad \widetilde{\mathfrak{Z}}^{\mathcal{N}=4,\textrm{short}}_{\frac{w}{2}}(a,q)=
\prod^{\infty}_{n=\frac{w}{2}}\prod^{\frac{w}{2}}_{m=-\frac{w}{2}} 
 \frac{1}{(1-a^m q^n)}\\
&
\cdot \prod^{\infty}_{n=\frac{w+1}{2}}\prod^{\frac{w-1}{2}}_{m=-\frac{w-1}{2}} 
 (1- a^m q^n)^2
\cdot \prod^{\infty}_{n=\frac{w}{2}+1}\prod^{\frac{w}{2}-1}_{m=-\frac{w}{2}+1} 
 \frac{1}{(1-a^m q^n)} \,.
\end{aligned}
\end{equation}
Taking the product of $\widetilde{\mathfrak{Z}}^{\mathcal{N}=4,\textrm{short}}_{\frac{w}{2}}(a,q)$ from all the spectrally-flowed sectors, we then have the full index of the vacuum representation of the chiral algebra: 
\begin{equation}\label{eq:FullindexWS}
\begin{aligned}
\widetilde{\mathfrak{Z}}^{\textrm{ws, Schur}}_{U(N)}(a,q)
= \prod^{\infty}_{w=1} \widetilde{\mathfrak{Z}}^{\mathcal{N}=4,\textrm{short}}_{\frac{w}{2}}(a,q)\,.
\end{aligned}
\end{equation}
Now that since the sign structures of the numerators and the denominators in \eqref{eq:FullIndexWSw} are the same, there is a huge amount of cancellation, which among other things removes the higher $n$ terms in \eqref{eq:FullIndexWSw}, and we end up with 
\begin{equation}\label{eq:FullIndexWScan}
\begin{aligned}
\widetilde{\mathfrak{Z}}^{\textrm{ws, Schur}}_{U(N)}(a,q)
= \prod^{\infty}_{w=1} \widetilde{\mathfrak{Z}}^{\mathcal{N}=4,\textrm{short}}_{\frac{w}{2}}(a,q)
= \prod^{\infty}_{w=1}\frac{(1-q^w)}{(1-a^{\frac{w}{2}} q^{\frac{w}{2}})(1-a^{-\frac{w}{2}} q^{\frac{w}{2}})}\,.
\end{aligned}
\end{equation}
It is then straightforward to check that it matches the spacetime result $\IFf^{\textrm{Schur},U(N)}(a,q)$ in  \eqref{eq:IndexLargeNSchurUN}:
\begin{equation}\label{eq:MatchFullSchurIndex}
\begin{aligned}
\widetilde{\mathfrak{Z}}^{\textrm{ws, Schur}}_{U(N)}(a,q)
=\prod^{\infty}_{k=1}\frac{1}{1-\IFsl^{\textrm{Schur}}_{\textrm{vec}}(a^k,q^k)}
=\IFf^{\textrm{Schur},U(N)}(a,q)\,.
\end{aligned}
\end{equation}

\subsection{Relation to compactification-independent subsector of \texorpdfstring{\ASTT}{AdS3S3M4} and comparison to twisted holography}
\label{ssec:AST}

The Schur subsector of the worldsheet theory of \ASF\ should capture the worldsheet description of \AST, and hence should be the ``compactification-independent" subsector of the worldsheet theory of the \ASTT\ case. 

In fact, the analysis is greatly facilitated by the fact that \cite{Gaberdiel:2021jrv} has already shown that if one also imposes the analogue of the ``wedge-mode" constraint on the \\
\ASTT\ case, one lands on a subsector of \ASTT\ case that is independent of the excitations of $T^4$. 
Then since before imposing the physical state condition, the worldsheet Schur subsector of the \ASF\ worldsheet CFT has the same field contents as the worldsheet CFT of \ASTT, namely $4$ symplectic bosons plus $2$ complex fermions, and imposing the physical state condition cuts down the spectrum drastically and restricts to only the  compactification-independent part of the \ASTT\ worldsheet CFT, we conclude that the Schur subsector of the worldsheet theory of \ASF\  captures an \AST\ subsector of \ASF.

\medskip

This is also consistent with the twisted holography of \cite{Costello:2018zrm}. 
Recall that  the bulk dual of the full 4D $\mathcal{N}=4$ SYM is the IIB string in \ASF\ whereas the bulk dual of the chiral algebra subsector is the B-model in the group manifold SL$(2,\mathbb{C})$ (which is \AST) \cite{Costello:2018zrm}.
The conjectured worldsheet dual of the full 4D theory is the  \ASF\ worldsheet CFT given by \cite{Gaberdiel:2021qbb,Gaberdiel:2021jrv}. The result of the current paper suggests that the worldsheet dual of the chiral algebra should be the BPS sector of the compactification-independent part of the \ASTT\ worldsheet theory, after appropriate twisting.\footnote{It would be very interesting to elucidate this connection further, in particular, the roles played by the different flux backgrounds and the details of the twisting from the worldsheet perspective.}  
Since the resulting worldsheet theory can only access the \AST\ part of the full \ASTT\ geometry, this is then consistent with the fact that the bulk dual of the chiral algebra lives in the group manifold SL$(2,\mathbb{C})\sim$ \AST\ $\subset$ \ASF\ background. 

Note that this does not mean that the worldsheet theory that corresponds to the chiral algebra is the B-model worldsheet theory on the nose. In fact, it is the A-model counterpart that sits more naturally as a subsector of the \ASTT\ worldsheet theory.\footnote{We thank Edward Mazenc for pointing this out to us.} However, in the presence of $\mathcal{N}=4$ supersymmetry, the A-model and B-model are related via hyper-K\"ahler rotation and can thus be considered equivalent.

\subsection{A comparison between chiral algebra \texorpdfstring{$\mathcal{W}_{\infty}$}{Winfinity} and higher-spin \texorpdfstring{$\mathcal{W}_{\infty}$}{Winfinity}}
\label{ssec:chiralVShigherspin}

Before we end this section, we compare this chiral algebra with the $\mathcal{N}=4$ $\mathcal{W}_{\infty}$ algebra that is the symmetry algebra of the boundary CFT of a Vasiliev higher spin gravity. 
To distinguish these two $\mathcal{N}=4$ $\mathcal{W}_{\infty}$ algebras, we call the former ``chiral algebra $\mathcal{W}_{\infty}$" and the latter ``higher spin $\mathcal{W}_{\infty}$".
The higher spin $\mathcal{W}_{\infty}$ algebra consists of one field per spin for $s=2,3,\dots,\infty$, and its character is \cite[eq.\ (2.32)]{Gaberdiel:2017dbk}
\begin{equation}\label{eq:N4characterExp}
\ChNf_{\mathcal{W}_{\infty}}(a,q)= \ChNf_{\rm{vac}} (a,q)  + \sum_{h=1}^{\infty}\ChLoNf_{h,0} (a,q) \,,
\end{equation}
where $\ChNf_{{\rm{vac}}} (a,q)$ is the vacuum character of the ${\cal N}=4$ superconformal algebra \eqref{eq:N4chaVac},
and the $\ChLoNf_{h} (a,q)$ are the ${\cal N}=4$ superconformal characters of the long multiplets whose bottom components have spin $0$, see \eqref{eq:N4chaLong}.

\medskip

Comparing the expansion \eqref{eq:ZUNsingleBPSExp} with \eqref{eq:BPSindexh} vs.\ \eqref{eq:N4characterExp} with \eqref{eq:N4chaVac} and \eqref{eq:N4chaLong}, the difference between these two $\mathcal{W}$ algebras is immediate. 
First of all, the chiral algebra $\mathcal{W}_{\infty}$ has one generating multiplet for each spin $s=\frac{n}{2}$, where $n\in \mathbb{N}$, whereas the higher spin $\mathcal{W}_{\infty}$ has one generating multiplet for each spin $s=n$.
More importantly, in the  chiral algebra $\mathcal{W}_{\infty}$, all the generating fields are BPS, whereas in the higher spin $\mathcal{W}_{\infty}$ algebra all but the $\mathcal{N}=4$ superconformal generators are non-BPS.
Therefore, the chiral algebra is independent of the coupling, whereas the higher spin algebra is the symmetry (sub)algebra only at zero-coupling. 
Namely, although the higher  spin $\mathcal{W}_{\infty}$ is a subalgebra of the dual CFT, all the symmetries generated by the long multiplets in \eqref{eq:N4characterExp} (which all contain  higher spin fields, even for $h=1$) disappear once one moves away from the free point.
The result is that, at generic coupling, we have only the $\mathcal{N}=4$ superconformal symmetry, which manifests itself as the symmetry algebra of the dual CFT of \ASTM, as confirmed by a computation of the anomalous dimensions as the string coupling is turned on \cite{Gaberdiel:2015uca}.

\section{Summary and discussion}

In this paper, we have derived the chiral algebra of 4D $\mathcal{N}=4$ SYM from its conjectured worldsheet CFT. 
We first extracted the subsector of the worldsheet theory that captures the Schur operators of the 4D theory.
This worldsheet ``Schur subsector" consists of precisely half of the free fields of the full worldsheet CFT, namely $4$ symplectic bosons plus $2$ complex fermions out of the $8$ symplectic bosons plus $4$ complex fermions. 
It generates a very large $\mathcal{N}=4$ $\mathcal{W}_{\infty}$ algebra (which we called the worldsheet Schur algebra), consisting of both short and long multiplets of the 2D $\mathcal{N}=4$ superconformal algebra. 

\medskip

Next, using the fact that the chiral algebra is independent of the coupling and hence should only be generated by short multiplets of the 2D $\mathcal{N}=4$ superconformal algebra, we obtain the chiral algebra by removing all the long multiplets. 
The resulting chiral algebra is an $\mathcal{N}=4$ $\mathcal{W}_{\infty}$ algebra, generated by all the short $\mathcal{N}=4$ multiplets $\ChShNf_{h=\frac{w}{2}}(a,q)$, with $h=s$ running through all positive integers and positive half-integers.
This is consistent with the picture that at the free point, for which the worldsheet theory is valid, the symmetry is vastly enhanced, captured by the bigger characters $\pfWSw^{(w)}(a,q)$.
Away from the free point, most of the spectrum gets lifted and the remaining algebra is generated by $\ChShNf_{h=\frac{w}{2}}(a,q)$.
We have checked that the chiral algebra agrees with the result from the SYM computation. 

\medskip

Finally, we have also shown that the worldsheet Schur algebra corresponds to the compactification-independent part of \ASTM.
This is from the observation that before imposing the physical state condition, the worldsheet Schur subsector of the \ASF\ worldsheet CFT has the same field contents as the worldsheet CFT of \ASTT, namely $4$ symplectic bosons plus $2$ complex fermions.
Imposing the physical state condition cuts down the spectrum drastically and restricts to only the  compactification-independent part of the \ASTM\ worldsheet CFT.
Roughly speaking, the Schur subsector captures an \AST\ subsector of \ASF.

\medskip

The result of this paper thus gives a further check for the proposed worldsheet CFT dual for 4D $\mathcal{N}=4$ SYM of \cite{Gaberdiel:2021qbb,Gaberdiel:2021jrv}, and it also connects to the worldsheet CFT of the free symmetric orbifold of $T^4$.
Finally, we comment that, since the computation is done at the level of the spectrum, it is not sensitive to the origin or the meaning of the conjectured physical state condition of \cite{Gaberdiel:2021qbb,Gaberdiel:2021jrv} itself.

\bigskip

Before we end this paper, we list some interesting problems for future research.
\begin{itemize} 
\item 
The analysis of this paper can be straightforwardly generalized to the 4D $\mathcal{N}=2$ superconformal theories obtained from orbifolding 4D $\mathcal{N}=4$ SYM. 
Instead of AdS$_5$$\times$S$_5$, the bulk geometry is AdS$_5$$\times$SE$_5$ where SE$_5$ are those Sasaki-Einstein manifolds that are the tip of the Calabi-Yau cone $\mathbb{C}^2/\mathbb{Z}_n\times \mathcal{C}$, and their worldsheet duals are given in \cite{Gaberdiel:2022iot}.
Having a full family of such worldsheet chiral algebras would highlight the connection between the chiral algebra and the corresponding geometry. 	
	 	
%\item A natural problem is to reproduce the  correlation functions of the chiral algebra directly from the \ASF\ worldsheet theory. 
%For this, we need to have a better understanding of the physical state condition from first principles.
%Alternatively, one can also try to reproduce these correlation functions from the (BPS sector of the) ``compactification-independent" part of the \ASTT\ worldsheet theory or from the corresponding subsector in the symmeric orbifold CFT.
%This would also help us to better understand the origin of the physical state condition of the \ASF\ worldsheet theory, since both the \ASTT\ worldsheet theory and the symmetric orbifold CFT are fairly well-understood. 

\item It would be interesting to consider the ``compactification-dependent" subsector 
of \ASTT\ worldsheet theory or that 
of the symmetric orbifold CFT, not just the BPS part, and then perform a  conformal perturbation computation\footnote{Note that this computation would require the knowledge of chiral algebra correlation functions.} to move the theory away from the free point to see directly how the long multiplets in \eqref{eq:difference} get lifted.

\item It would be interesting to study the worldsheet counterparts of the BPS sectors with fewer supersymmetries of 4D $\mathcal{N}=4$ SYM. 
Even in the large-$N$ limit, there are interesting features such as the indices of different saddles \cite{Cabo-Bizet:2018ehj,Choi:2018hmj}, SL$(3,\mathbb{Z})$ modularity \cite{Goldstein:2020yvj,Jejjala:2021hlt,Jejjala:2022lrm}, the Bethe Ansatz equations \cite{Benini:2018mlo,Benini:2018ywd} etc. 
One might try to see whether the worldsheet theory can shed any new light on these problems. 

%\item Finally, it would be very satisfying to be able to perform the worldsheet analogue of the twisting procedure of \cite{Beem:2013sza} that extracts the chiral algebra. 

\end{itemize}

\section*{Acknowledgements}

We would like to thank Edward Mazenc, Matthias Gaberdiel, and Wolfgang Lerche for helpful discussions.
This work is supported by NSFC No.\ 11875064, No.\ 12275334, No.\ 11947302, and the Max-Planck Partnergruppen fund.
We are also grateful for the support and hospitality of the Max Planck Institute for Gravitational Physics at Potsdam, the Kavli Institute for Theoretical Physics at Santa Barbara, and the Issac Newton Institute for Mathematical Sciences at Cambridge University (during the Program ``Black holes: bridges between number theory and holographic quantum information" with EPSRC Grant ER/R014604/1) where part of this work was carried out. 

%\nocite{*}
\bibliographystyle{utphys} 
\bibliography{biblio}

\end{document}